\documentstyle[12pt,epsfig]{article}

\newcommand{\nc}{\newcommand}
\nc{\rnc}{\renewcommand}
\nc{\acs}{\arraycolsep}
\nc{\mc}{\multicolumn}
\nc{\bsk}{\baselineskip}
\nc{\vsp}{\vspace}
\nc{\hsp}{\hspace}
\nc{\stl}{\setlength}
\nc{\stc}{\setcounter}
\nc{\addl}{\addtolength}
\nc{\beq}{\begin{equation}}
\nc{\eeq}{\end{equation}}
\nc{\beqa}{\begin{eqnarray}}
\nc{\eeqa}{\end{eqnarray}}
\nc{\tfrac}[2]{\raisebox{.4ex}{\tiny $\frac{#1}{#2}$}}
\nc{\romlist}{ \setcounter{num1}{0}%
  \begin{list}{(\roman{num1})}{\usecounter{num1}} }
\nc{\arblist}{ \setcounter{num1}{0}%
  \begin{list}{(\arabic{num1})}{\usecounter{num1}} }
\nc{\alphlist}{ \setcounter{num2}{0}%
  \begin{list}{(\alph{num2})}{\usecounter{num2}} }
\nc{\bullist}{\begin{list}{$\bullet$}{ }}
\nc{\nr}{\\ \hline}
\nc{\hrl}{{\center \stl{\unitlength}{\textwidth}
 \begin{picture}(1,0)  \put(0,0){\line(1,0){1}}
 \end{picture} \vsp{.001\bsk} }}
\nc{\cents}{{\scriptsize$\mbox{\rm C}\!\!\!\mbox{\raisebox{.2ex}%
{$|$}}\,\,\,\,$}}
\nc{\figsp}[5]{\begin{figure}[#1] \vsp{#2} \caption[#4]{#3}
\label{#5} \vsp{2\bsk} \end{figure}}
\nc{\fig}{\figsp{tbp}}
\nc{\figb}{\figsp{b}}
\nc{\figh}{\figsp{h}}
\nc{\llist}{\begin{list}{}{} \stl{\labelsep}{.4in}}
\nc{\lit}[2]{
 \item[\raggedright #1]{#2}}
\nc{\lbit}[2]{
 \item[\raggedright\bf #1]{#2}}
\nc{\lemit}[2]{
 \item[\raggedright\em #1]{#2}}
\nc{\lbemit}[2]{
 \item[\raggedright\bf\em #1]{#2}}
\nc{\clst}[1]{\stl{\coltwo}{\textwidth}
\addl{\coltwo}{-#1} \addl{\coltwo}{-5.56ex} \newline
\begin{tabular}{p{#1}p{\coltwo}} \citem{}{}}

\nc{\citem}[2]{{\raggedright \bf #1} & #2 \\ }
\nc{\cemitem}[2]{{\raggedright \em #1} & #2 \\ }
\nc{\cbemitem}[2]{{\raggedright \bf \em #1} & #2 \\ }
\nc{\cend}{\citem{}{} \end{tabular}
\mbox{}
}
\nc{\SSP}{{\rm \hsp{.4in}}}
\nc{\SSPP}{{\rm \hsp{.2in}}}
\nc{\ds}{\displaystyle}
\nc{\tx}{\textstyle}
\nc{\scst}{\scriptstyle}
\nc{\sscst}{\scriptscriptstyle}
\nc{\prt}{\partial}
\nc{\fr}{\frac}
\nc{\lf}{\left}
\nc{\rt}{\right}
\nc{\la}{\langle}
\nc{\ra}{\rangle}
\nc{\V}{\vec}
\nc{\str}{\stackrel}
\nc{\ovl}{\overline}
\nc{\ul}{\underline}
\nc{\ovb}{\overbrace}
\nc{\ub}{\underbrace}
\nc{\wh}{\widehat}
\nc{\B}{\bar}
\nc{\D}{\dot}
\nc{\C}{\cdot}
\nc{\dd}{\ddot}
\nc{\tl}{\tilde}
\nc{\ha}{\hat}
\nc{\nn}{\nonumber}
\nc{\app}{\approx}
\nc{\al}{\alpha}
\nc{\RA}{\rightarrow}
\nc{\LRA}{\leftrightarrow}
\nc{\SRA}{\SSP\rightarrow\SSP}
\nc{\SSRA}{\SSPP\rightarrow\SSPP}
\nc{\dg}{\dagger}
\nc{\vp}{\varphi}
\nc{\ve}{\varepsilon}
\nc{\Dl}{\Delta}
\nc{\dl}{\delta}
\nc{\gm}{\gamma}
\nc{\Gm}{\Gamma}
\nc{\ep}{\epsilon}
\nc{\sg}{\sigma}
\nc{\Sg}{\Sigma}
\nc{\ua}{\uparrow}
\nc{\da}{\downarrow}
\nc{\lam}{\lambda}
\nc{\eql}[1]{\parbox{#1\textwidth}}
\nc{\eqm}[1]{\makebox[#1\textwidth][l]}
\nc{\enu}[1]{\mbox{\hspace{.4in}(\theequation.#1)}}
\nc{\son}{\\ \\ \ds}
\nc{\stw}{\\ & \\ \ds}		%   Equation formatting   %
\nc{\sth}{\\ & & \\ \ds}
\nc{\sfo}{\\ & & & \\ \ds}
\nc{\sfi}{\\ & & & & \\ \ds}
\nc{\A}{& \ds}
\nc{\bbr}{\lf\{\rule[-1.5ex]{0in}{0.01in}\rt.}
\nc{\hf}{\fr{1}{2}}
\nc{\mhf}{\mbox{\footnotesize$\hf$}}
\nc{\dv}{\/!}
\nc{\dint}{\int\!\!\int}
\nc{\tint}{\int\!\!\dint}               % integrals %
\nc{\qint}{\int\!\!\tint}
\nc{\Pd}[2]{\fr{\prt #1}{\prt #2}}
\nc{\Pdt}[1]{\Pd{#1}{t}}
\nc{\Pdx}[1]{\Pd{#1}{x}}
\nc{\Pdy}[1]{\Pd{#1}{y}}
\nc{\Pdz}[1]{\Pd{#1}{z}}           	%  Derivatives  %
\nc{\Pdr}[1]{\Pd{#1}{r}}
\nc{\Pds}[1]{\Pd{#1}{s}}
\nc{\Dv}[2]{\fr{d#1}{d#2}}
\nc{\Dvt}[1]{\Dv{#1}{t}}
\nc{\Dvx}[1]{\Dv{#1}{x}}
\nc{\Dvy}[1]{\Dv{#1}{y}}
\nc{\Dvz}[1]{\Dv{#1}{z}}
\nc{\Dvr}[1]{\Dv{#1}{r}}
\nc{\Drs}[1]{\Dv{#1}{s}}
\nc{\inpp}[3]{\la #1| #2| #3\ra}
\nc{\inp}[2]{\inpp{#1}{#2}{#1}}
\nc{\rb}[1]{| #1\ra}
\nc{\lb}[1]{\la#1|}
\nc{\dtpp}[2]{\lb{#1}\rb{#2}}		%  Inner Products  %
\nc{\dtp}[1]{\dtpp{#1}{#1}}
\nc{\otpp}[2]{\rb{#1}\lb{#2}}
\nc{\otp}[1]{\otpp{#1}{#1}}
\rnc{\L}{{\cal L}}                      %  Miscellaneous  %
\nc{\lapp}{\mbox{\raisebox{-.6ex}{$\,\stackrel{\textstyle <}{\sim}\,$}}}
\nc{\gapp}{\mbox{\raisebox{-.6ex}{$\,\stackrel{\textstyle >}{\sim}\,$}}}
\nc{\gpx}{g_1^p(x,Q^2)}
\nc{\gpz}{g_1^p(z,Q^2)}
\nc{\muq}{\lf(\fr{\mu^2}{Q^2}\rt)}
\nc{\xy}{(\fr{x}{y})}
\nc{\ASQ}{\al_s(Q^2)}
\nc{\dqx}{\Dl q_i(x,Q^2)} \nc{\dqy}{\Dl q_i(y,Q^2)}
\nc{\dQ}{\Dl q_i(Q^2)}
\nc{\dgx}{\Dl g(x,Q^2)} \nc{\dgy}{\Dl g(y,Q^2)}
\nc{\dG}{\Dl g(Q^2)}
\nc{\xq}{(x,Q^2)} \nc{\yq}{(y,Q^2)}
\nc{\Tt}{\tl{t}} \nc{\Ts}{\tl{s}} \nc{\Tu}{\tl{u}}
\nc{\Hs}{\ha{s}} \nc{\Ht}{\ha{t}} \nc{\Hu}{\ha{u}}
\nc{\Hsg}{\hat{\sg}}
\nc{\GeV}{\mbox{\rm GeV}}
\nc{\pS}{\!\not{\!p}}  \nc{\kS}{\!\not{\!k}}
\nc{\poS}{\!\not{\!p}_1}  \nc{\pwS}{\!\not{\!p}_2}
\nc{\ptS}{\!\not{\!p}_3}  \nc{\pfS}{\!\not{\!p}_4}
\nc{\AS}{\!\not{\!\!A}}  \nc{\ASS}{\!\not{\!\!A}^*}
\nc{\BS}{\!\not{\!\!B}}  \nc{\BSS}{\!\not{\!\!B}^*}
\nc{\Tr}{\mbox{\rm Tr}}
\nc{\pT}{p_T}
\nc{\xT}{x_T}
\nc{\AoS}{\!\not{\!\!A}_1}  \nc{\AoSS}{\!\not{\!\!A}_1^*}
\nc{\AwS}{\!\not{\!\!A}_2}  \nc{\AwSS}{\!\not{\!\!A}_2^*}
\nc{\BoS}{\!\not{\!\!B}_1}  \nc{\BoSS}{\!\not{\!\!B}_1^*}
\nc{\BwS}{\!\not{\!\!B}_2}  \nc{\BwSS}{\!\not{\!\!B}_2^*}
\nc{\aS}{\!\not{\!a}}  \nc{\bS}{\!\not{\!b}}
\nc{\aoS}{\!\not{\!a}_1}  \nc{\boS}{\!\not{\!b}_1}
\nc{\awS}{\!\not{\!a}_2}  \nc{\bwS}{\!\not{\!b}_2}
\nc{\anS}{\!\not{\!a}_n}  \nc{\bnS}{\!\not{\!b}_n}

\begin{document}

\pagestyle{empty}

\hfill {\bf McGill--95/2}

\hfill {\bf hep-ph/9503489}

\hfill January 1995

\vspace{1cm}

\begin{center} \begin{Large} \begin{bf}
Heavy-Quark Production by Polarized and \\
Unpolarized Photons in Next-to-Leading Order
\end{bf} \end{Large} \end{center}
\vglue 0.35cm
{\begin{center}
 B.\ Kamal,$^{a,1}$
Z.\ Merebashvili$^{a,2,*}$
and A.P.\ Contogouris$^{a,b,3}$\end{center}}
\parbox{6.4in}{\leftskip=1.0pc
{\it a.\ Department of Physics, McGill University, Montreal,
Qc., H3A 2T8, Canada}\\
%\vglue -0.25cm
%\baselineskip=14pt
{\it b.\ Nuclear and Particle Physics, University of Athens,
Athens 15771, Greece}
}
\begin{center}
\vglue 1.0cm
\begin{bf} ABSTRACT \end{bf}
\end{center}
%\vglue 1.0cm
{%\rightskip=1.5pc
 %\leftskip=1.5pc
 %\tenrm\baselineskip=12pt
 \noindent
Complete analytical results for the production of heavy-quark
pairs by polarized and unpolarized photons in next-to-leading
order are presented. Two-, three- and
two plus three-jet cross sections for
total photon spin $J_z=0,\pm 2$ are presented for $b\bar{b}(g)$
production. The 2-jet cross sections are considered as
a background to $\gamma\gamma\rightarrow H^*\rightarrow b\bar{b}$
(standard model).
Top production, not
too far above threshold, is also considered for $J_z=0,\pm 2$.
For both $b$- and $t$-quark production, the higher order
QCD corrections are found to be significant.
}

\vspace{1in}
\begin{center}
(Phys.\ Rev.\ D {\bf 51}, 4808 (1995))
\end{center}

\renewcommand{\thefootnote}{\arabic{footnote}}
\addtocounter{footnote}{1}
\footnotetext{e-mail: kamal@wind.phy.bnl.gov}
\addtocounter{footnote}{1}
\footnotetext{e-mail: zaza@hep.physics.mcgill.ca}
\addtocounter{footnote}{1}
\footnotetext{e-mail: apcont@hep.physics.mcgill.ca,
acontog@atlas.uoa.ariadne-t.gr}

\newpage

\pagestyle{plain}
\setcounter{page}{1}

%\vglue .3cm
\begin{center}\begin{large}\begin{bf}
I. INTRODUCTION
\end{bf}\end{large}\end{center}
\vglue .3cm

Higher (next-to-leading) order corrections (HOC) for heavy-quark
($Q$,$\overline{Q}$) production in unpolarized particle collisions
have
been determined in detail.$^{\,\mbox{\scriptsize \ref{Nason},
\ref{Drees}}}$
For polarized
particle collisions, however, analytical results were
still absent. Even for the unpolarized case, only virtual + soft
corrections have been presented analytically.$^{\,\mbox{\scriptsize
\ref{Been}}}$
Apart from general reasons, well known from unpolarized
reactions, knowledge of HOC for $Q$, $\overline{Q}$ production in
polarized
processes is important for several special reasons.

Beginning with polarized $\gamma\gamma$ collisions, which is the subject
of the present work, one reason of special interest is the following.
A $\gamma\gamma$ collider becomes particularly important for searches
of the standard model Higgs boson when its mass is below the $W^+W^-$
threshold. Then the predominant decay is $H\rightarrow b\bar{b}$ and the
background comes from $\gamma\gamma\rightarrow b\bar{b}$ with direct or
resolved
photons. Leaving  aside the latter, for the moment, use of
polarized photons of equal helicity (when the angular momentum
has $J_z=0$) suppresses this background by a factor
$m_b^2/s$.$^{\,\mbox{\scriptsize \ref{Hab},\ref{Borden}}}$
This holds, however, only for the
lowest order of $\alpha_s$. HOC necessarily involve the subprocess
$\gamma\gamma\rightarrow b\bar{b} g$, and gluon emission permits the
$b\bar{b}$ system
to have $J\neq 0$ without suppression;
this may result in a sizable background.
Of course, another reason
the $J_z=0$ channel is important is that the Higgs
signal comes entirely from it. Thus, we maximize the
Higgs to background ratio in two different ways.

Furthermore, at higher energies, it will be possible to produce
top-quarks
in photon-photon collisions. This, when combined with other
data on top-quark production from $e^+e^-$ and $p\bar{p}$ collisions,
should certainly improve our knowledge of the top-quark parameters.
The HOC could have a significant effect on the threshold behaviour.
It is also interesting to examine the spin dependence of the
HOC in this region.

In this paper we present complete analytical results for heavy
quark production by both polarized and unpolarized photons.
Numerical results are presented for 2-, 3- and 2+3-jet cross
sections for the cases where the initial photons have total
spin $J_z=0$ and $J_z=\pm 2$. For $b$-quark production, this is
analyzed as a background to Higgs production. We also consider
$t$-quark production for energies not too far above threshold.

The analytical results presented here
are also useful in determining
the production
of heavy quarks in polarized
photon-proton (proton-proton) collisions. This is because
the process $\gamma\gamma\rightarrow Q\overline{Q}(g)$ is the
Abelian (QED)
part of the
subprocess $\gamma g\rightarrow Q\overline{Q}(g)$
($gg\rightarrow Q\overline{Q}(g)$),
which is by far the dominant subprocess
in $\gamma$-$p$ ($p$-$p$)
collisions.$^{\,\mbox{\scriptsize \ref{Nason},\ref{Been}}}$
The non-Abelian part
of $\vec{\gamma}\vec{g}\rightarrow Q\overline{Q}(g)$
($\vec{g}\vec{g}\rightarrow Q\overline{Q}(g)$)
remains to be calculated.

\vglue 1cm
\begin{center}\begin{large}\begin{bf}
II.  LEADING ORDER CROSS SECTIONS
\end{bf}\end{large}\end{center}
\vglue .3cm

\begin{figure}
\vsp{-.5in}
\vspace{-4cm}
\noindent\hsp{0.75in}
\epsfig{figure=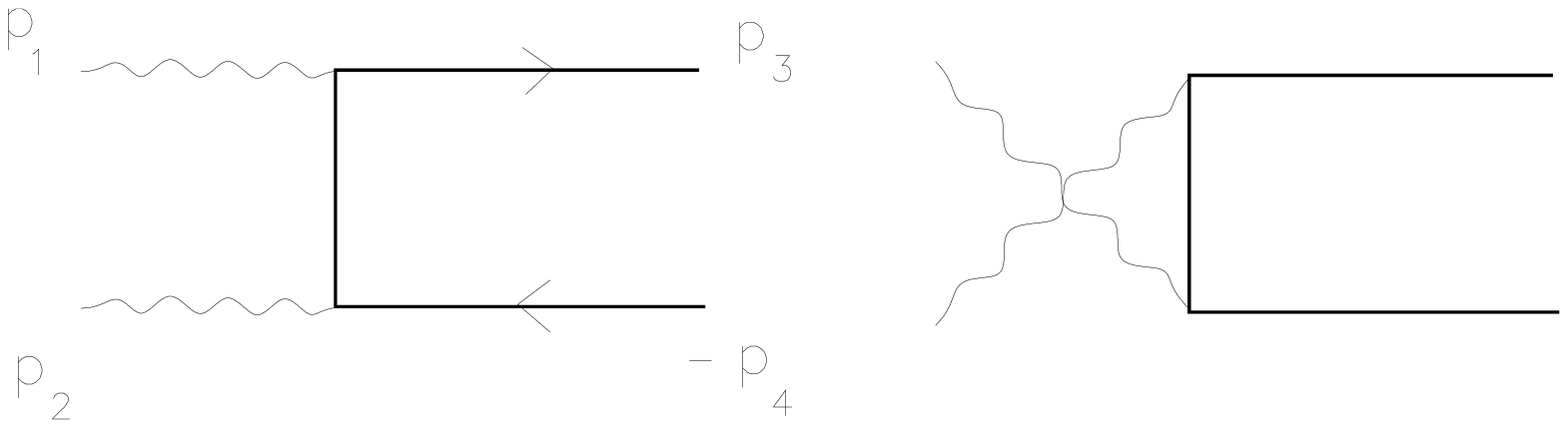,width=6.5in,height=4in,rheight=0in,rwidth=0in}
\vspace{6.5cm}
\vsp{.5in}
\caption{ \protect\small\baselineskip=12pt
Lowest order contributions to $\gm\gm\RA Q\ovl{Q}$.
}
\end{figure}

The contributing graphs are shown in Fig.\ 1. We introduce the
variables (momenta as in figure)
\begin{equation}
s \equiv (p_1+p_2)^2, {\rm \hspace{.2in}} t \equiv T-m^2 \equiv
(p_1-p_3)^2 - m^2,
{\rm \hspace{.2in}} u \equiv U-m^2 \equiv (p_2-p_3)^2 -m^2
\end{equation}
and
\begin{equation}
s_2 \equiv S_2 - m^2 \equiv (p_1+p_2-p_3)^2 - m^2 = s+t+u.
\end{equation}
where $m$ is the heavy-quark mass.
Defining
\begin{equation}
v \equiv 1+\frac{t}{s}, {\rm \hspace{.2in}} w \equiv \frac{-u}{s+t}
\end{equation}
we may express
\begin{equation}
\label{tus2}
t = -s(1-v), {\rm \hspace{.2in}} u = -svw, {\rm \hspace{.2in}}
 s_2 = sv(1-w).
\end{equation}
The polarized and unpolarized squared amplitudes are defined,
respectively, as
\begin{equation}
\label{msqr}
\Delta |M|^2 \equiv \frac{1}{2} (|M(+,+)|^2 - |M(+,-)|^2),
{\rm \hspace{.2in}}
|M|^2 \equiv \frac{1}{2} (|M(+,+)|^2 + |M(+,-)|^2),
\end{equation}
where $M(\lambda_1,\lambda_2)$ denotes the Feynman amplitude with
photons $p_1$, $p_2$ having helicity $\lambda_1$, $\lambda_2$
respectively. The same holds for the cross sections.

For the examples considered in this paper, it is of interest
to calculate (numerically) the cross sections for a specific helicity
state, $\sigma(\lambda_1,\lambda_2)$.
We present analytical results for the polarized and unpolarized
cross sections $\Delta \sigma$, $\sigma$. From (\ref{msqr}) we can
obtain the desired cross sections via
\begin{equation}
\label{sigma}
\sigma(+,+) = \sigma + \Delta \sigma, {\rm \hspace{.4in}}
\sigma(+,-) = \sigma - \Delta \sigma.
\end{equation}
Defining
\begin{equation}
K(\varepsilon) \equiv \frac{m^{-2\varepsilon}}{s}\frac{\pi
(4\pi)^{-2+\varepsilon}}{\Gamma(1-\varepsilon)}
\left(\frac{tu-sm^2}{sm^2}\right)^{-\varepsilon}
\end{equation}
we may express the $n$ ($=4-2\varepsilon$)-dimensional
2-body phase space as
\begin{equation}
[\Delta] \frac{d\sigma_{2\rightarrow 2}}{dvdw} =
K(\varepsilon) \left[ (2m)^2 [\Delta] |M|^2_{2\rightarrow 2}
\right] \delta(1-w).
\end{equation}
It will become necessary to work in $n$ dimensions when we
determine the HOC (see next section for details).

The resulting leading-order (LO) cross sections are, in $n$ dimensions,
\begin{eqnarray}
\label{LO}
\nonumber
\Delta \frac{d\sigma_{\rm LO}}{dvdw} &=& 32 \pi^2 K(\varepsilon)
\delta(1-w) N_C \alpha^2
e_{Q}^4 \mu^{4\varepsilon}  \left\{
{}-\frac{t^2+u^2}{tu} + 2\frac{sm^2}{tu}\left(\frac{s^2}{tu} - 2\right)
\right\} \\
\frac{d\sigma_{\rm LO}}{dvdw} &=& 32 \pi^2 K(\varepsilon)
\delta(1-w) N_C \alpha^2
e_{Q}^4 \mu^{4\varepsilon} \left\{
\frac{t^2+u^2}{tu} + 4\frac{sm^2}{tu} - 4\left(\frac{sm^2}{tu}\right)^2
\right\}
\end{eqnarray}
where $N_C$ (=3) is the number of quark colors and $e_{Q}$ is
the fractional charge of the heavy quark.
Making use of (\ref{tus2}), (\ref{sigma}) we see explicity that
$d\sigma_{\rm LO}(+,+)/dvdw$ is suppressed by order $m^2/s$.

\vglue 1cm
\begin{center}\begin{large}\begin{bf}
III.  LOOP CONTRIBUTIONS
\end{bf}\end{large}\end{center}
\vglue .3cm

\begin{figure}
\vsp{-.5in}
\vspace{-1.5cm}
\noindent\hsp{1in}
\epsfig{figure=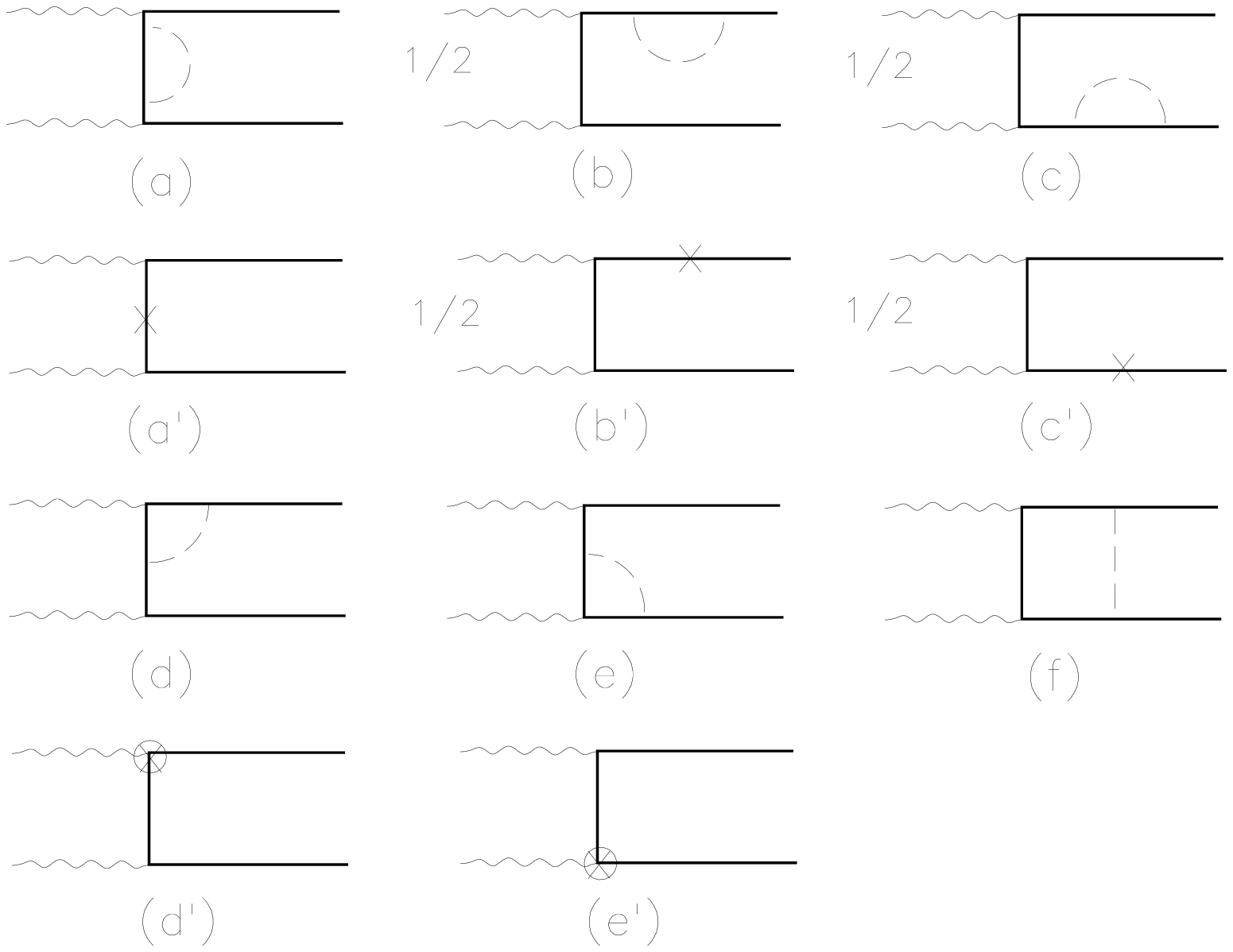,width=6.5in,height=5in,rheight=0in,rwidth=0in}
\vspace{11.5cm}
\vsp{.5in}
\caption{ \protect\small\baselineskip=12pt
Loop graphs for $\gm\gm \RA Q\ovl{Q}$.
{\bf (a)}--{\bf (c)} self-energy diagrams;
{\bf (a$'$)}--{\bf (c$'$)} mass counterterm diagrams corresponding
to the graphs (a)--(c); {\bf (d),(e)} vertex diagrams;
{\bf (d$'$),(e$'$)} dimensional reduction counterterm diagrams
corresponding to graphs (d),(e); {\bf (f)} box diagram.
}
\end{figure}

The loop contributions arise from the
diagrams of Fig.\ 2 and
their $p_1\leftrightarrow p_2$ interchange. These diagrams contain
both ultraviolet
and infrared singularities. To regularize them, we use dimensional
reduction$^{\,\mbox{\scriptsize \ref{Seigel}}}$,
where the momenta are in $n$ dimensions and everything
else is in four dimensions.
This facilitates the handling of the Levi-Civita tensor
$\varepsilon^{\mu\nu\lambda\rho}$.
As we will show below, the analytical expressions
for the cross sections are {\em regularization scheme independent}
once all the
contributions (including the gluonic bremsstrahlung) are added.
Throughout, we work in the Feynman gauge.

The heavy-quark mass and wave function renormalizations
are performed on-shell. The self-energy graphs are shown in Figs.\ 2
(a)--(c) and the corresponding mass counterterm diagrams in Figs.\ 2
(a$'$)--(c$'$). The factor $1/2$ multiplying (b)-(c$'$) comes from
wave function renormalization. The bare mass and wave function are
determined in terms of the renormalized ones via
\begin{equation}
m_0 = Z_m m_r, {\rm \hspace{.4in}} \Psi_0 = Z_2^{1/2} \Psi_r,
\end{equation}
where $Z_m$ and $Z_2$ are the mass and wave function renormalization
constants.
Define
\begin{equation}
C_\varepsilon \equiv \frac{\Gamma(1+\varepsilon)}{(4\pi)^2}
\left(\frac{4\pi\mu^2}{m^2}\right)^\varepsilon
\end{equation}
where $\mu$ is an arbitrary mass scale which enters via the
coupling in $n$ dimensions: $g\rightarrow g\mu^\varepsilon$. In
dimensional reduction
we find, to order $g^2$,
\begin{equation}
Z_m = 1-3g^2C_\varepsilon C_F \left(\frac{1}{\varepsilon'} + \frac{5}{3}
\right),
{\rm \hspace{.2in}} Z_2 = 1 - g^2 C_\varepsilon C_F \left(
\frac{1}{\varepsilon'} + 5 + \frac{2}{\varepsilon}\right)
\end{equation}
with $C_F = 4/3$. We use $1/\varepsilon'$ to indicate which terms are of
ultraviolet origin.

In dimensional reduction we must add to the vertex diagrams of
Figs.\ 2 (d), (e) appropriate counterterms (d$'$), (e$'$) in order
to satisfy the Ward identity$^{\,\mbox{\scriptsize \ref{ward}}}$
\begin{equation}
\label{Ward}
Z_1 = Z_2
\end{equation}
between vertex and self-energy graphs, with $Z_1$ denoting the
vertex renormalization constant. The Feynman rule for this vertex
counterterm is found to be (in $n$ dimensions)
\begin{equation}
\label{cterm}
\gamma^\mu \rightarrow \frac{-g^2}{(4\pi)^2} C_F \gamma_\varepsilon^\mu
\frac{1}{\varepsilon'}
\end{equation}
with
\begin{equation}
\gamma_\varepsilon^\mu = (g^{\mu\nu}-g^{\mu\nu}_n)\gamma_\nu, {\rm
\hspace{.4in}}
 g^{\mu\nu}_n g^n_{\mu\nu} = g^{\mu\nu}_n g_{\mu\nu} =n;
\end{equation}
here $g_n^{\mu\nu}$ represents the $n$-dimensional metric tensor
with, formally, $n<4$.

When all the contributions to the physical cross section
(including gluonic bremsstrah-lung) are added,
the result is free of infrared divergences as there are no
collinear singularities here. Thus the only scheme dependent part
might come from the   vertex and self-energy graphs. Having satisfied
the Ward identity (\ref{Ward}) though, means that the scheme
dependent part of the corrections
cancels between vertex and self-energy graphs. This
was explicitly verified by calculating the vertex and self-energy
graphs in dimensional regularization.
We also checked explicitly that there are no differences
between reduction and regularization arising from
any other contributions.
More specifically, to obtain the
dimensional regularization result for any particular contribution
given in this paper, simply replace the LO term by the corresponding
LO term from dimensional regularization.
When all the contributions are added, the scheme dependent part of the
LO term cancels along with the $1/\varepsilon$ infrared divergence
multiplying it.
Hence, the absence of
collinear divergences or vacuum polarization graphs leads to
scheme independence.

As was stated in Refs.\ \ref{DY} and \ref{LLbar}, a counterterm
like
(\ref{cterm}) was used to remove an unphysical term.
General one-loop
counterterms have been developed$^{\,\mbox{\scriptsize \ref{Korner}}}$ to
convert unpolarized dimensional reduction results into the corresponding
dimensional regularization results for the purely massless case.
Also, certain equivalences between dimensional reduction and
dimensional regularization have been
noted.$^{\,\mbox{\scriptsize \ref{Jones}}}$
In the present case however, satisfaction of (\ref{Ward}) is
sufficient to ensure scheme independence.

Adding the contributions of Figs.\ 2 (a)--(e$'$)
(and the $t\leftrightarrow u$ interchange) resulted in the
ultraviolet finite vertex plus self-energy cross section
\begin{eqnarray}
\label{vse}
%\nonumber
\lefteqn{ \frac{d\sigma_{\rm vse}}{dvdw}
= -\frac{16}{\varepsilon}\pi\alpha_s C_F C_{\varepsilon}
\frac{d\sigma_{\rm LO}}{dvdw}
 + \delta(1-w) \frac{CK(0)}{(4\pi)^2}
\{ 2 A_1 [4(\xi(2)-{\rm Li}_2(\frac{T}{m^2}))(1+3\frac{m^2}{t}) }  \\
\nonumber &-& \ln(-\frac{t}{m^2}) (8 -6\frac{t}{T}
-\frac{t^2}{T^2})- 2-\frac{t}{T}]
+A_2\ln(-\frac{t}{m^2})+A_3({\rm Li}_2(\frac{T}{m^2})-\xi(2))+A_4 +
 (t\leftrightarrow u) \}
\end{eqnarray}
where
\begin{equation}
C\equiv (4\pi)^3 C_F N_C \alpha_s \alpha^2 e_Q^4 \mu^{6\varepsilon}.
\end{equation}
The corresponding polarized cross section,
$\Delta d\sigma_{\rm vse}/dvdw$ can be obtained by replacing the $A_i$ and
$d\sigma_{\rm LO}/dvdw$ in
(\ref{vse}) by $\Delta A_i$ and $\Delta d\sigma_{\rm LO}/dvdw$, respectively.
The $[\Delta] A_i$ are given in Appendix B.
We will use this notation throughout.
We note the term $\sim 1/\varepsilon$ in (\ref{vse}) representing
an infrared divergence. Also, note that $[\Delta] A_1$
is proportional to the LO squared amplitude without the
$t\leftrightarrow u$ interchange (see Appendix B, Eq.\ \ref{A1LO}).

Since $[\Delta] d\sigma_{\rm LO}/dvdw$ is in general regularization
scheme dependent to ${\cal O}({\varepsilon})$ (working in $n$ dimensions),
we see explicitly that truly scheme independent cross sections will result
only when all contributions are added and all infrared divergences
are cancelled.

In order to evaluate the box graph of Fig.\ 2 (f), we must reduce
the resulting tensor integrals to scalar ones (conveniently
listed in Ref. \ref{Been}) using projective tensor
techniques.$^{\,\mbox{\scriptsize \ref{PV}}}$  The tensor integrals have
the general form
\begin{eqnarray}
\lefteqn{D^{0,\mu,\mu\nu,\mu\nu\lambda}(k_1,k_2,k_3,m_1,m_2,m_3,m_4)
\equiv} \\
\nonumber
& & \mu^{2\varepsilon} \int\!\! \frac{d^n q}{(2\pi)^n}
\frac{1,q^{\mu},q^\mu q^\nu,q^\mu q^\nu q^\lambda}
{(q-m_1)^2[(q+k_1)^2-m_2^2][(q+k_1+k_2)^2-m_3^2]
[(q+k_1+k_2+k_3)^2-m_4^2]}
\end{eqnarray}
where the $k_i$ are general momenta. As an example, the vector
box integral we encounter has the decomposition
\begin{equation}
D^\mu(p_4,-p_2,-p_1,0,m,m,m) = p_4^\mu D_{11} - p_2^\mu D_{12}
- p_1^\mu D_{13}.
\end{equation}
In general, the scalar coefficients $D_{ij}$ are not independent.
This simplifies somewhat the calculation. Noting that
\begin{equation}
D^\mu(p_4,-p_2,-p_1,0,m,m,m)
= - D^\mu(p_3,-p_1,-p_2,0,m,m,m),
\end{equation}
we obtain
\begin{equation}
D_{12} = D_{11} - D_{13},
\end{equation}
since the $D_{ij}$ in both integrals are the same, due to the
fact that they are scalars. Using the same approach, we reduce
the number of independent $D_{ij}$ from seven to five in
$D^{\mu\nu}$ and from thirteen to eight in $D^{\mu\nu\lambda}$.
This method was quite helpful in keeping the very large intermediate
expressions as short as possible.

Adding the contribution of Fig.\ 2 (f) and the $t\leftrightarrow u$
interchange
gives the virtual box cross section
%\newpage
\begin{eqnarray}
\label{box}
\nonumber  \frac{d\sigma_{\rm box}}{dvdw}
&=& 16\pi\alpha_s C_F C_{\varepsilon} \frac{d\sigma_{\rm LO}}{dvdw}
  \frac{2m^2-s}{s\beta}\{2\ln(x)[\frac{1}{2\varepsilon}
       - \ln(\beta)] + 2{\rm Li}_2(-x) - 2{\rm Li}_2(x)-3\xi(2)\} \\
\nonumber
&+& \delta(1-w) \frac{CK(0)}{(4\pi)^2}
\{- 8B_1\frac{2m^2-s}{s \beta} \ln(x) \ln(-t/m^2)
+ 2\frac{B_2}{\beta}[\ln(x)
         (4\ln(1+x)-\ln(x) \\ \nonumber
&-&4\ln(-t/m^2)) + 4{\rm Li}_2(-x)+2\xi(2)]+
         2B_3\ln^2(x)+4\frac{B_4}{\beta}\ln(x)+4B_5\ln(-t/m^2) \\
{}  &+& 8B_6{\rm Li}_2(T/m^2)+4B_7\xi(2)+4B_8 + (t\leftrightarrow u)\}
\end{eqnarray}
where
\begin{equation}
\xi(2) = \frac{\pi^2}{6}, {\rm \hspace{.4in}}
\beta \equiv \sqrt{1-4m^2/s}, {\rm \hspace{.4in}} x \equiv
\frac{1-\beta}{1+\beta}.
\end{equation}
The $[\Delta] B_i$ are given in Appendix B. We see again the
infrared divergence $\sim 1/\varepsilon$.

Independent calculations were performed using
FORM$^{\,\mbox{\scriptsize \ref{Verm}}}$ and
REDUCE$^{\,\mbox{\scriptsize \ref{Hearn}}}$.
The latter proved useful in factoring the
expressions and cancelling powers in the denominators.

\vglue 1cm
\begin{center}\begin{large}\begin{bf}
IV.  GLUONIC BREMSSTRAHLUNG CONTRIBUTIONS
\end{bf}\end{large}\end{center}
\vglue .3cm

The bremsstrahlung diagrams are shown in Fig.\ 3. Squaring these
diagrams (plus their $p_1\leftrightarrow p_2$ interchange),
we obtain the $2\rightarrow 3$ particle squared amplitude
\begin{eqnarray}
\label{M23}
\nonumber \lefteqn{(2m)^2  |M|^2_{2\rightarrow 3}  =} \\ \nonumber
& & C[\frac{\tilde{e}_1}{s_2^2} + e_2\,\,p_2\cdot k
+ \frac{e_3}{p_2\cdot p_4}
+ \frac{e_4}{p_2\cdot p_4^{\,\,2}}
+ e_5\frac{p_2\cdot k}{p_1\cdot p_4^{\,\,2}}
{}+ \frac{\tilde{e}_6}{s_2\,\,p_3\cdot k} + \frac{\tilde{e}_7}{p_3\cdot
k^{\,\,2}} \\ \nonumber
&+& \frac{e_8}{p_1\cdot p_4\,\, p_2\cdot p_4}
+ e_9\frac{p_2\cdot k^{\,\,2}}{p_1\cdot p_4}
+ e_{10}\frac{p_2\cdot k}{p_1\cdot p_4}
+ \frac{\tilde{e}_{11}/s_2}{p_2\cdot p_4\,\, p_3\cdot k}
+ \frac{\tilde{e}_{12}}{p_2\cdot p_4\,\, p_3\cdot k^{\,\,2}} \\
&+& \frac{e_{13}}{p_2\cdot p_4^{\,\,2}\,\, p_3\cdot k}
{}+ \frac{\tilde{e}_{14}}{p_2\cdot p_4^{\,\,2}\,\, p_3\cdot k^{\,\,2}}
+ e_{15}\frac{p_2\cdot k^{\,\,2}}{p_3\cdot k}
+ e_{16}\frac{p_2\cdot k}{p_3\cdot k}
] + (p_1\leftrightarrow p_2, t\leftrightarrow u)
\end{eqnarray}
As before, we may obtain $\Delta |M|^2_{2\rightarrow 3}$ by replacing the
$e_i$ in (\ref{M23}) by $\Delta e_i$.
The $[\Delta] e_i$ are given in Appendix B. Again, independent
calculations were performed using FORM and REDUCE. The former
proved useful in partial fractioning and other reductions of the
dot-products.

To obtain the total bremsstrahlung contribution to $[\Delta]
d\sigma/dvdw$, we perform the phase-space integrals in the frame
where $p_4$ and $k$ are back-to-back. We find (in agreement
with Ref.\ \ref{Been}) for the $2\rightarrow 3$ phase space
\begin{equation}
\label{twothreeps}
[\Delta] \frac{d\sigma_{\rm Br}}{dvdw} =
K(\varepsilon) \frac{C_\varepsilon}{\mu^{2\varepsilon}}\tilde{f}
(\varepsilon)
\int\!\!
d\Omega  (2m)^2 [\Delta] |M|_{2\rightarrow 3}^2
\end{equation}
where
\begin{equation}
\tilde{f}(\varepsilon) \equiv \frac{(m^2)^{1-\varepsilon}}
{S_2^{1-\varepsilon}}
\frac{sv}{2\pi} \left( \frac{sv}{m^2} \right)^{1-2\varepsilon}
(1-w)^{1-2\varepsilon}
\end{equation}
and
\begin{equation}
\label{angint}
\int\!\! d\Omega \equiv \int_0^\pi \!\! d\theta_1
\sin^{1-2\varepsilon}\theta_1 \int_0^\pi\!\!  d\theta_2
\sin^{-2\varepsilon}\theta_2.
\end{equation}
The gluon angles $\theta_1$ and $\theta_2$ are defined in
Appendix A along with all the momenta.

We first evaluate all the phase space integrals in four dimensions
since, for $w\neq 1$, all the integrals are finite.
For $w=1$, the
terms in (\ref{M23}) with coefficients $\tilde{e}_i$ are singular
through the relation$^{\,\mbox{\scriptsize \ref{EllFurm}}}$
\begin{equation}
(1-w)^{-1-2\varepsilon} = -\frac{1}{2\varepsilon}\delta(1-w)
+ \frac{1}{(1-w)_+}
+ {\cal O}(\varepsilon)
\end{equation}
where the function $1/(1-w)_+$ is defined through
\begin{equation}
\int_{w_1}^1 \!\! dw \frac{f(w)}{(1-w)_+} =
\int_{w_1}^1 \!\! dw \frac{f(w)-f(1)}{(1-w)} + f(1)\ln(1-w_1).
\end{equation}
This means that, for these terms, the integrals must also be
evaluated in $n$ dimensions in the limit $w\rightarrow 1$,
keeping their ${\cal O}(\varepsilon)$ part.
The resulting integrals are straightforward.

The final result is, with $\bar{y}\equiv \sqrt{(t+u)^2-4m^2s}$,
\begin{eqnarray}
\label{br}
\nonumber \lefteqn{ \frac{d\sigma_{\rm Br}}{dvdw} =} \\ \nonumber
& & \frac{C K(0)}{(4\pi)^2}  2\pi\tilde{f}(0)
\{ \frac{s_2(s+u)}{4S_2} e_2 + \frac{2S_2}{s_2(s+u)}
\ln\frac{S_2}{m^2}\,\, e_3
+  \frac{4S_2}{m^2(s+u)^2} e_4 \\ \nonumber
&+& e_5 I_5 + e_8 I_8 + e_9 I_9 + e_{10} I_{10} + e_{13} I_{13}
+ e_{15} I_{15} +  e_{16} I_{16} + (t\leftrightarrow u) \} \\ \nonumber
&+& \frac{1}{(1-w)_+} \frac{C K(0)}{(4\pi)^2} \frac{1}{S_2}
\{  \tilde{e}_1 + \frac{2S_2}{\bar{y}}
\ln\frac{T+U-\bar{y}}{T+U+\bar{y}} \,\, \tilde{e}_6
+ 4\frac{S_2}{m^2} \tilde{e}_7
+ s_2 I_{11} \tilde{e}_{11} \\
&+& s_2^2 I_{12} \tilde{e}_{12}
+ s_2^2 I_{14} \tilde{e}_{14} + (t\leftrightarrow u) \}
+ 8\pi\alpha_s C_F C_{\varepsilon} \frac{d\sigma_{\rm LO}}{dvdw}
\frac{1}{s\beta}
 \{(2m^2-s) \\ \nonumber
&\times& \left[2\ln x(2 \ln \frac{sv}{m^2}
-\frac{1}{\varepsilon}) -1 -2\left({\rm Li}_2\left(
\frac{-4\beta}{(1-\beta)^2}\right)
+ \ln^2 x\right)\right]
+ 2s\beta\left[1-2\ln\frac{sv}{m^2} + \frac{1}{\varepsilon}\right]\},
\end{eqnarray}
where the integrals $I_i$ are given in Appendix C.

\vglue 1cm
\begin{center}\begin{large}\begin{bf}
V.  PHYSICAL CROSS SECTIONS
\end{bf}\end{large}\end{center}
\vglue .3cm

\begin{figure}
\vsp{-.5in}
\vspace{-4cm}
\noindent\hsp{0.75in}
\epsfig{figure=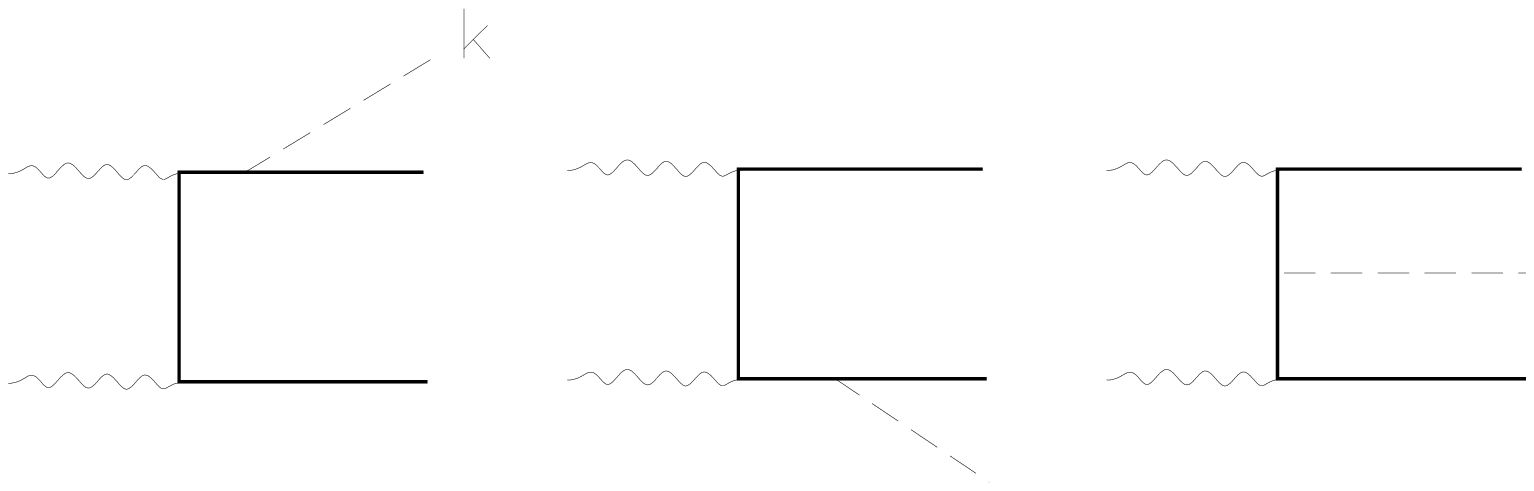,width=6.5in,height=4in,rheight=0in,rwidth=0in}
\vspace{7cm}
\vsp{.5in}
\caption{ \protect\small\baselineskip=12pt
Gluonic Bremsstrahlung graphs for $\gm\gm\RA Q\ovl{Q}g$.
}
\end{figure}

We may obtain the 2+3-jet cross section by adding (\ref{LO}),
(\ref{vse}), (\ref{box}),  and (\ref{br}):
\begin{equation}
[\Delta] \frac{d\sigma_{2+3}}{dvdw} =
[\Delta] \frac{d\sigma_{\rm LO}}{dvdw} + [\Delta]
\frac{d\sigma_{\rm vse}}{dvdw}
+ [\Delta] \frac{d\sigma_{\rm box}}{dvdw} + [\Delta]
\frac{d\sigma_{\rm Br}}{dvdw}
\end{equation}
We notice the cancellation of all the $1/\varepsilon$ infrared
divergences,
leading to a finite, scheme independent result.

At this point it is useful to note that for $s\gg 4m^2$, the LO
cross sections (\ref{LO}) are large in the forward and backward
directions.
Since jets going down the beam pipe are difficult to measure
experimentally, angular cuts are necessary for $b\bar{b}$ production
well above threshold.
At the same time, we reduce the $b\bar{b}$ background to the
Higgs signal.
This also helps eliminate
{\em resolved photon} contributions where the partons within the photon
participate, as opposed to the {\em direct} contributions,
which we present, where the photon
is structureless.
This is discussed at the end of this section.

Let $\theta_3$ denote the  angle between $p_3$ and
$p_1$ in the $\gamma\gamma$ c.m.
Then the integrated 2+3-jet cross section, with the constraint
$|\cos\theta_3|<\cos\theta_c$,
for some $\theta_c$, is given by
\begin{equation}
[\Delta] \sigma_{2+3}(s) = \int_{v_1}^{v_2}\!\!dv\int_{w_1}^1\!\!dw\,\,
\Theta(\cos^2\theta_c-\cos^2\theta_3) [\Delta] \frac{d\sigma_{2+3}}{dvdw}
\end{equation}
where
\begin{equation}
v_1 = \frac{1}{2}(1-\beta), {\rm \hspace{.2in}} v_2=\frac{1}{2} (1+\beta),
{\rm \hspace{.2in}}
w_1(v) = \frac{m^2}{s} \frac{1}{v(1-v)}
\end{equation}
and
\begin{equation}
\cos\theta_3 =
\frac{-(1-v-vw)}{\sqrt{(1-v+vw)^2-4m^2/s}}.
\end{equation}
Alternatively, we may convert to $d\sigma/d\!\cos\theta_3 dw$ and
integrate directly over $\theta_3$ and $w$.

The integrated 3-jet cross section is given by
\begin{eqnarray}
\nonumber
[\Delta] \sigma_3(s) &=& \frac{K(0)}{(4\pi)^2}
\int_{v_1}^{v_2}\!\!dv\int_{w_1}^1\!\!dw\,\,
\Theta(\cos^2\theta_c-\cos^2\theta_3) \tilde{f}(0)
\int\!\!d\Omega (2m)^2 [\Delta] |M|^2_{2\rightarrow 3} \\
\nonumber &\times& \Theta((p_3+k)^2-y_{\rm cut} s)
\Theta((p_4+k)^2-y_{\rm cut} s) \\
\nonumber
 &=& \frac{K(0)}{(4\pi)^2}
\int_{v_1}^{v_2}\!\!dv\int_{w_1}^{w_2}\!\!dw\,\,
\Theta(\cos^2\theta_c-\cos^2\theta_3) \tilde{f}(0)
\int\!\!d\Omega (2m)^2 [\Delta] |M|^2_{2\rightarrow 3} \\
&\times& \Theta((p_3+k)^2-y_{\rm cut} s);
{\rm \hspace{.4in}} w_2 = 1 - \frac{(y_{\rm cut} - m^2/s)}{v}.
\end{eqnarray}
The angular integral is given by (\ref{angint}) with $\varepsilon=0$.
The dot-products involved may be explicitly expressed as functions  of
$v$, $w$, $\theta_1$ and $\theta_2$ using the parametrizations of
Appendix A and Eqs.\ (\ref{tus2}).
We have imposed the constraints, $(p_3+k)^2>y_{\rm cut}s$
and $(p_4+k)^2>y_{\rm cut}s$. With a suitable choice of $y_{\rm cut}$
we may simultaneously cut out events with 2-jet topology and avoid
the soft divergence. We effectively eliminate the soft and collinear
gluons from the 3-jet cross section, with the degree of softness
and collinearity being specified by $y_{\rm cut}s$.

The desired 2-jet cross section is obtained by the difference
\begin{equation}
\label{twojet}
[\Delta] \sigma_2(s) = [\Delta] \sigma_{2+3}(s) - [\Delta] \sigma_3(s).
\end{equation}
Since $\sigma_{2+3}$ and $\sigma_3$ are both infrared finite and
separately observable quantities, this serves as a reliable
and unambiguous method for defining $\sigma_2$.

In discussing the numerical results, it will be convenient to
split $[\Delta] \sigma_{2+3}$ as follows,
\begin{equation}
[\Delta] \sigma_{2+3} = [\Delta] \sigma_{\rm LO} + [\Delta] \sigma_{\rm S}
 + [\Delta] \sigma_{\rm H},
\end{equation}
where $[\Delta] \sigma_{\rm S}$ represents the contribution to the HOC
coming from terms proportional to $\delta(1-w)$ and $1/(1-w)_+$,
and $[\Delta] \sigma_{\rm H}$ represents the rest.
In usual terminology,
$[\Delta] \sigma_{\rm S}$ represents virtual and soft contributions whereas
$[\Delta] \sigma_{\rm H}$ represents hard radiation.

So far we have only considered direct contributions, i.e.\ no
resolved photon contributions. The reason is the following.
Well above the $Q\overline{Q}$ threshold,
 $\sigma_{2+3}$ and $\sigma_3$ will certainly receive sizable
resolved photon contributions. Now, resolved photon events are
generally accompanied by a jet making small angles with respect
to the beam axis. For the 2-jet cross section (which
is of physical interest), experiment can reject resolved photon
events (and other unwanted events)
as being those for which the observed jets have total energy
measurably lower than $\sqrt{s}$.$^{\,\mbox{\scriptsize \ref{Borden}}}$
This is because, due to the angular cuts, experiment will not
observe the jet making small angles. Hence, there will be
missing energy.
Of course, we are assuming a
rather well defined initial photon energy, which may be experimentally
difficult.

For top-quark production, not too far above threshold, the resolved
contributions will be negligible in all the cross sections.
This is because the dominant resolved contribution comes from
$g \gamma \rightarrow Q\overline{Q}$, where the gluon originates from one
of the
initial photons, having a fraction $x$ of its momentum.
Near threshold, the gluon will have to carry a large fraction
of the photon's momentum;
and for $x\rightarrow 1$, the  gluon distribution in
the photon is highly suppressed. As well, 3-jet states arising
from hard gluonic radiation will be suppressed due to the
restricted phase space. The (near) absence of resolved
contributions and the non-suppression of the $J_z=0$ cross section
for $2\rightarrow 2$ kinematics, not too far above threshold,
implies that we needn't worry about
whether the events are 2- or 3-jet (even though 3-jet events are
either very seldom or none, depending on $s$).

\vglue 1cm
\begin{center}\begin{large}\begin{bf}
VI.  NUMERICAL RESULTS
\end{bf}\end{large}\end{center}
\vglue .3cm

Here we present numerical results for $b$- and $t$-quark production
in next-to-leading order.
Throughout, we evaluate $\alpha_s(Q^2)$ (2-loop)
with $Q^2=s$, $\Lambda=0.2\,\,\mbox{\rm GeV}$
and the number of flavors taken as $N_F=5$ since we are well
above the $b\bar{b}$ threshold. We take $m_b=4.7\,\,\mbox{\rm GeV}$ and
$m_t=174\,\,\mbox{\rm GeV}$.$^{\,\mbox{\scriptsize \ref{Abe}}}$
For 3-jet cross sections, we use $y_{\rm cut}=
0.15$. Some justification for this choice of $y_{\rm cut}$ is in
order. Experimentally, it is useful to have a small value of
$y_{\rm cut}$ so that for the 2-jet cross section we eliminate,
as much as possible, events with 3-jet topology via (\ref{twojet}).
Theoretically, there are limitations. If one chooses $y_{\rm cut}$
too small, then the infrared divergence ruins the perturbation
expansion, since the 3-jet cross section becomes unphysically
large. To control this, an all-orders resummation would be
required. We find that $y_{\rm cut}=0.15$ is the most suitable
choice in light of the above considerations.

Fig.\ 4(a) presents $\sigma_{\rm LO}(+,+)$, $\sigma_{2+3}(+,+)$,
$\sigma_3(+,+)$ and $\sigma_2(+,+)$ for $b$-quark production in the
range $20<\sqrt{s}<200\,\,\mbox{\rm GeV}$ with $\theta_c=30^0$. As
expected,
the LO cross section is highly suppressed for large $\sqrt{s}$,
but not the 3-jet. In fact $\sigma_3(+,+)$ makes a sizable contribution
to $\sigma_{2+3}(+,+)$. Hence $\sigma_2(+,+)$ gets somewhat suppressed
relative to $\sigma_{2+3}(+,+)$. For $20\mbox{
\raisebox{-.6ex}{$\,\stackrel{\textstyle <}{\sim}\,$}}
\sqrt{s}\mbox{\raisebox{-.6ex}{$\,\stackrel{\textstyle <}{\sim}\,$}}
 40\,\,\mbox{\rm GeV}$ the
corrections $\sigma_{2+3}-\sigma_{\rm LO}$ are seen to be slightly
negative.

\begin{figure}
\vsp{-.6in}
\hspace{1.35in}\noindent
\epsfig{figure=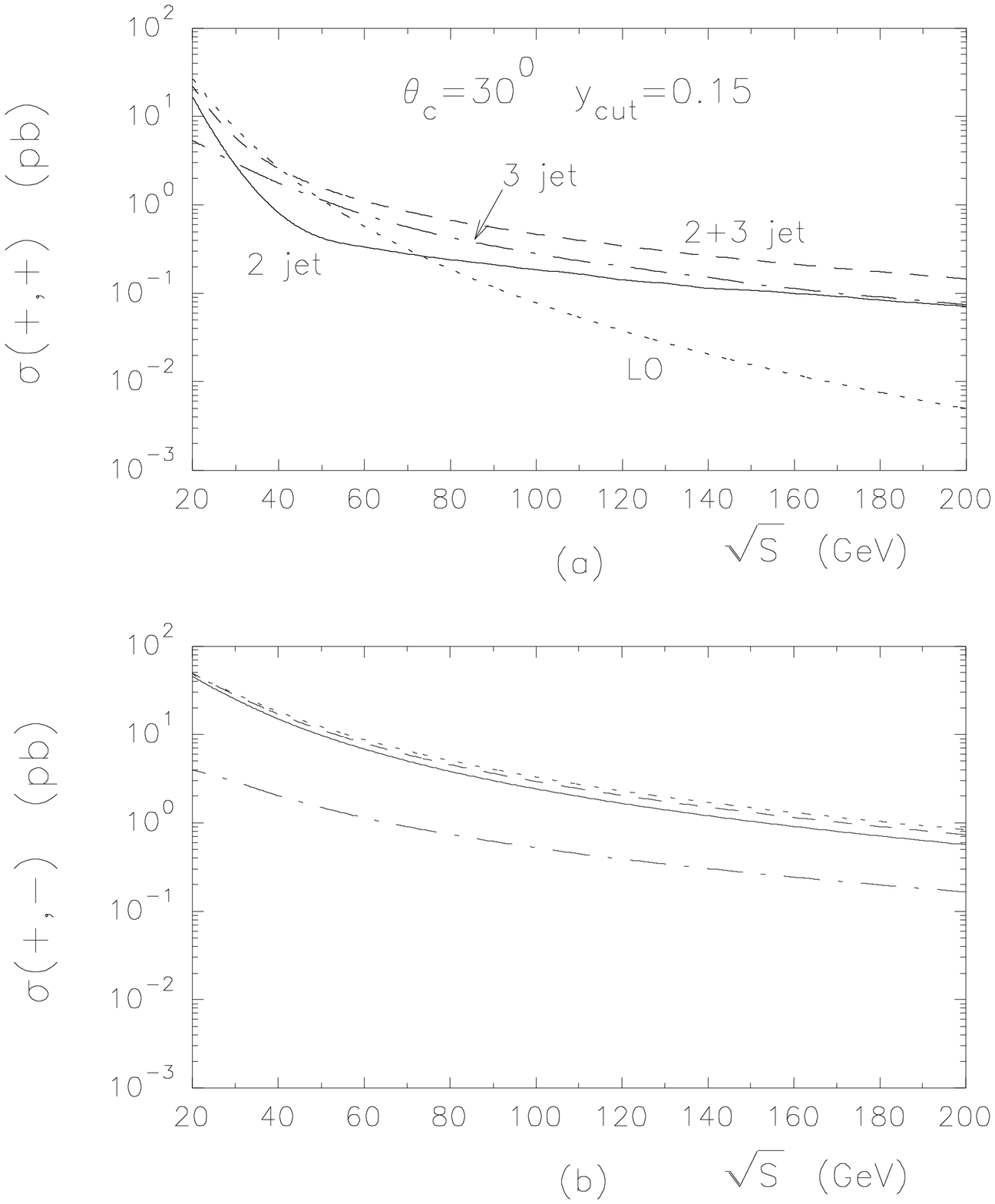,width=3.5in,height=5.1in,rheight=0in,rwidth=0in}
\vspace{5.6in}
\vspace{-.2in}
\caption{ \protect\small\baselineskip=12pt
Cross sections for $\gm\gm \RA b\B{b}(g)$:
$\sg_{\rm LO}$ (dotted line), $\sg_{2+3}$ (dashed),
$\sg_3$ (dash-dotted) and $\sg_2$ (solid), with
$\theta_c=30^0$ and $y_{\rm cut}=0.15$ for
$20<\protect\sqrt{s}<200\,\,\GeV$; {\bf (a)} $\sg(+,+)$; {\bf (b)}
$\sg(+,-)$.
}
\end{figure}
Fig.\ 4(b) presents the same cross sections for $J_z=\pm 2$, i.e.\
$\sigma_{\rm LO}(+,-)$, $\sigma_{2+3}(+,-)$,
$\sigma_3(+,-)$ and $\sigma_2(+,-)$. The major difference is that
$\sigma_{\rm LO}(+,-)$ and $\sigma_{2+3}(+,-)$ suffer no suppression
at large $\sqrt{s}$. Hence the 3-jet contribution to
$\sigma_{2+3}(+,-)$ is not so significant and $\sigma_2(+,-)$
remains large. We also notice, $\sigma_{2+3}\mbox{
\raisebox{-.6ex}{$\,\stackrel{\textstyle <}{\sim}\,$}}\sigma_{\rm LO}$
throughout.

\begin{figure}
\vsp{-.6in}
\hspace{1.35in}\noindent
\epsfig{figure=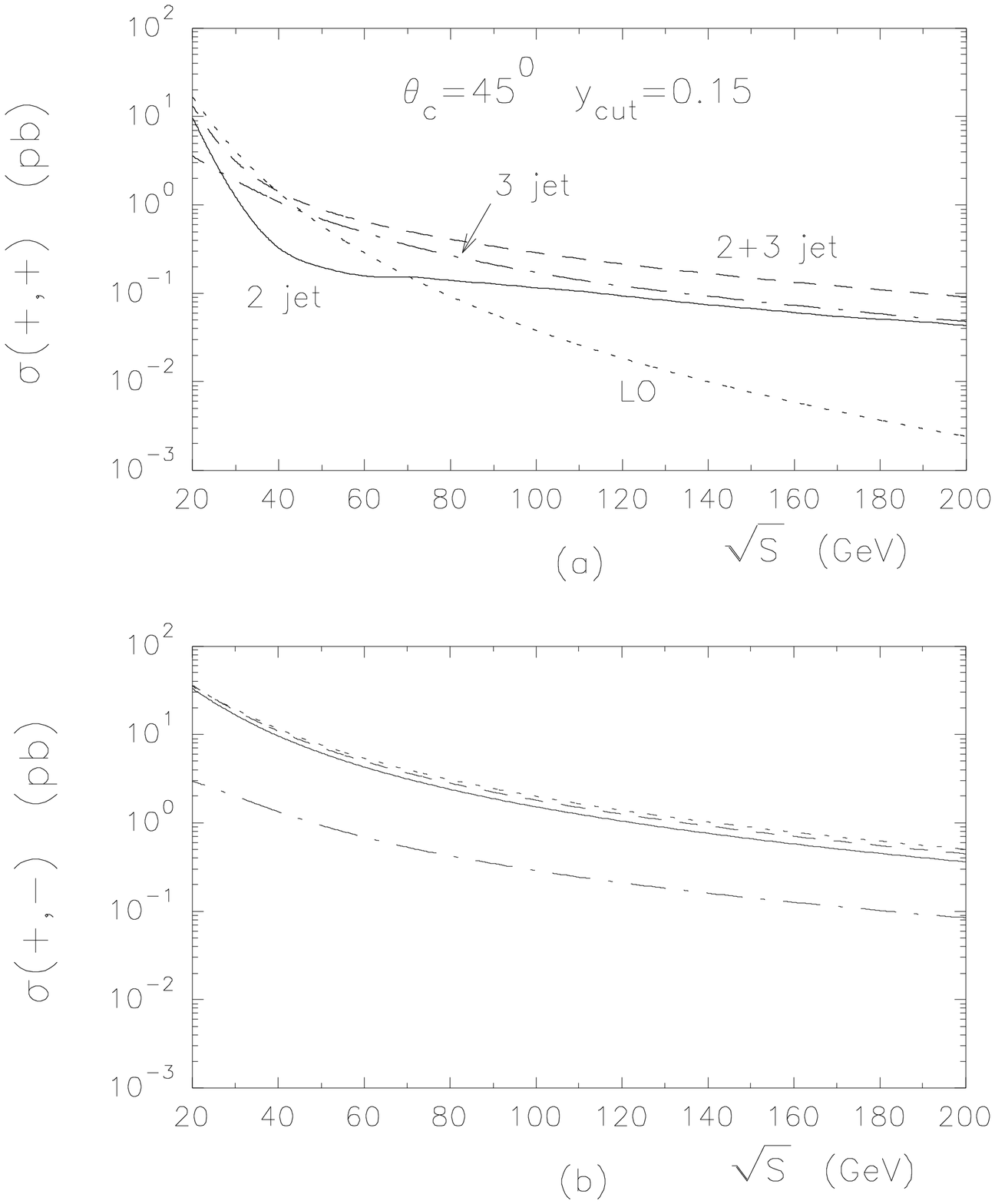,width=3.5in,height=5.1in,rheight=0in,rwidth=0in}
\vspace{5.6in}
\vspace{-.2in}
\caption{ \protect\small\baselineskip=12pt
Same as Fig.\ 4, except with $\theta_c=45^0$.
}
\end{figure}
Figs.\ 5 (a),(b) present the same quantities as in Figs.\ 4
(a),(b) except with $\theta_c=45^0$. The major difference is that
the cross sections are smaller everywhere and $\sigma_2(+,+)$ is
particularly suppressed for $30\mbox{\raisebox{-.6ex}
{$\,\stackrel{\textstyle <}{\sim}\,$}}\sqrt{s}
\mbox{\raisebox{-.6ex}{$\,\stackrel{\textstyle <}{\sim}\,$}}
 60\,\,\mbox{\rm GeV}$.
This reflects the fact that the 2-jet events tend to
occur at smaller angles.

An interesting feature of the HOC arises for both $\sigma_{2+3}$
and $\Delta\sigma_{2+3}$. In both cases, $\sigma_{\rm S}$ and
$\sigma_{\rm H}$ are much larger that $\sigma_{\rm LO}$, for
$s\gg 4m^2$. But they have opposite sign and are of almost
equal magnitude, leading to large cancellations. In other words,
the ``virtual + soft'' part conspires with the ``hard'' part
to yield HOC which are under control.

\begin{figure}
\hspace{1.25in}\noindent
\epsfig{figure=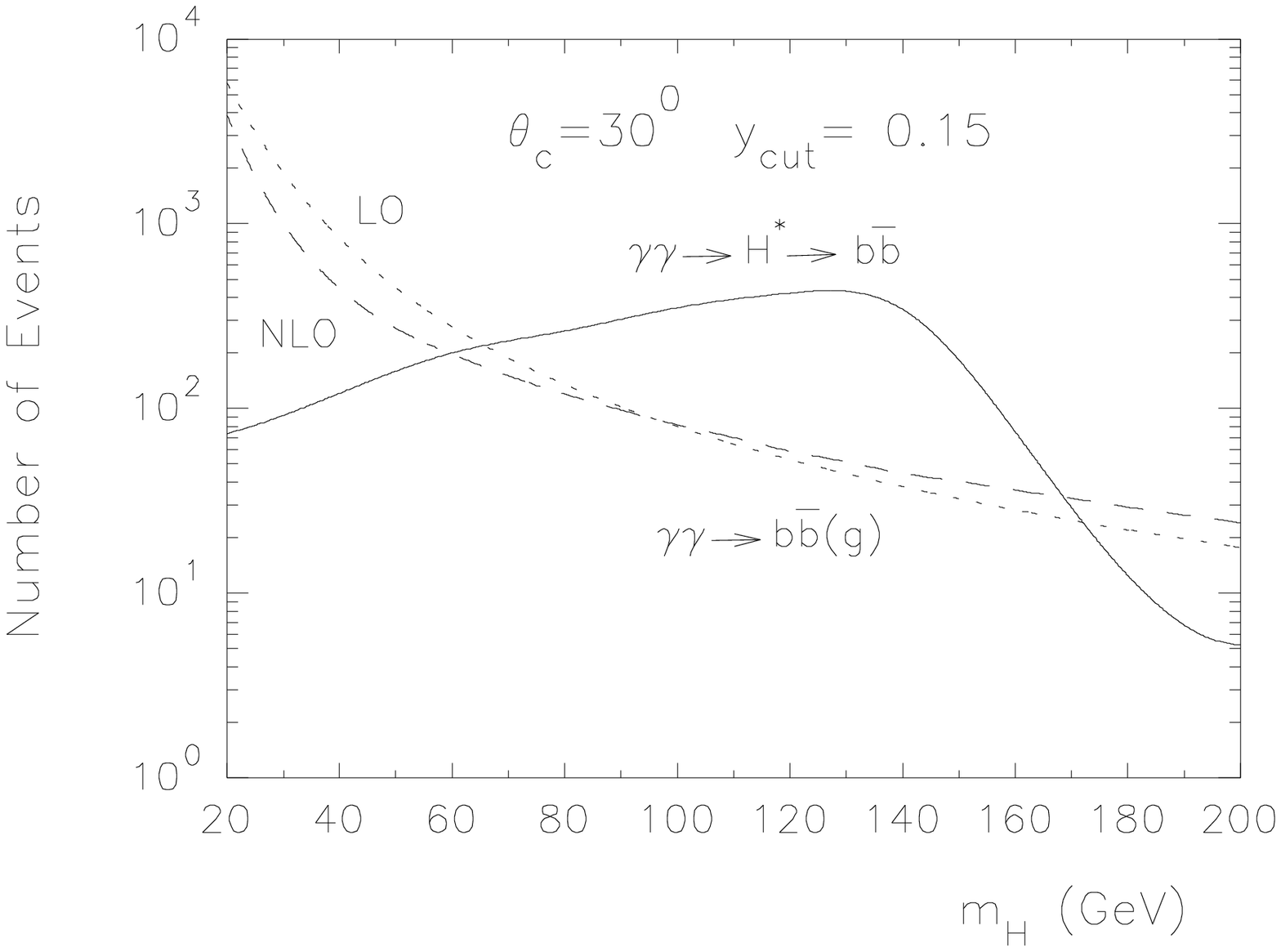,width=3.6in,height=3.6in,rheight=0in,rwidth=0in}
\vspace{5.6in}
\vspace{-1.8in}
\caption{ \protect\small\baselineskip=12pt
Two-jet $b\B{b}$ background to standard model
Higgs decay: $\gm\gm\RA H^*\RA
b\protect\B{b}$ (solid line), $\sg_{\rm LO}$ (dotted) and
$\sg_2$ (dashed) for $20<m_{\protect\rm H}<200\,\,\GeV$. Number of
Higgs events taken from Ref.\ \protect\ref{Hab}.
Here $\theta_c=30^0$,
$<\lam_1\lam_2>=0.8\,$. The other experimental parameters are
described in the text.
}
\end{figure}
Fig.\ 6 presents the 2-jet background to the Higgs decay
{}$\gamma\gamma\rightarrow H^* \rightarrow b\bar{b}$. We have used the
standard model Higgs cross section
of Ref.\ \ref{Hab} which takes $\theta_c=30^0$  and an
average value of $<\lambda_1\lambda_2>=0.8$. The photons are produced by
laser backscattering off electrons
(positrons) at an $e^+e^-$ collider with $E_{e^+e^-}=500\,\,
\mbox{\rm GeV}$.
As well, Ref.\ \ref{Hab} uses an
effective integrated luminosity of $L_{\rm eff}=20\,\,{\rm fb}^{-1}$
and a $\gamma\gamma$ energy spread of $\Gamma_{\rm expt}=5\,\,
\mbox{\rm GeV}$;
$\sqrt{s} = m_{\rm H} \pm \Gamma_{\rm expt}/2$.
Using the expression of Ref.\ \ref{Hab} for converting the
$\gamma\gamma\rightarrow b\bar{b}(g)$ cross section into number of
 events, we
obtain the LO and 2-jet curves shown in Fig.\ 6.

At large $\sqrt{s}$, the increase in $\sigma_2(+,+)$ relative
to $\sigma_{\rm LO}(+,+)$ is compensated by a decrease in
$\sigma_2(+,-)$ relative to $\sigma_{\rm LO}(+,-)$, so that
$\sigma_2(<\lambda_1\lambda_2>=0.8)$ doesn't change radically.
In the end,
the 2-jet cross section is still well below the Higgs signal
for $90\mbox{\raisebox{-.6ex}{$\,\stackrel{\textstyle <}{\sim}\,$}}
{}m_{\rm H}\mbox{\raisebox{-.6ex}{$\,\stackrel{\textstyle <}{\sim}\,$}}
150\,\,\mbox{\rm GeV}$. With higher degrees of
polarization, we could do even better.

\begin{figure}
\vsp{-.6in}
\hspace{1.35in}\noindent
\epsfig{figure=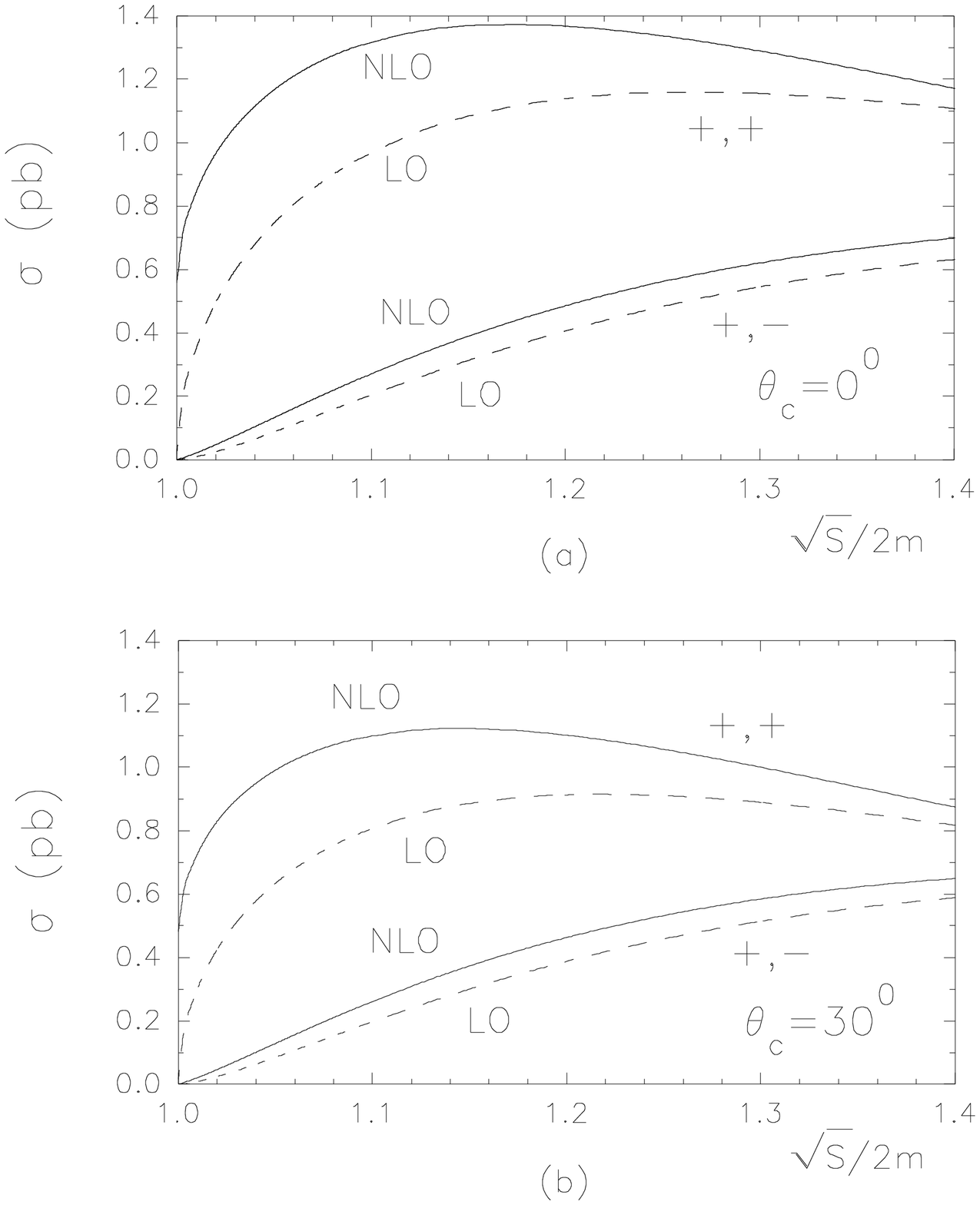,width=3.5in,height=5.1in,rheight=0in,rwidth=0in}
\vspace{5.6in}
\vspace{-.2in}
\caption{ \protect\small\baselineskip=12pt
Cross sections for $\gm\gm\RA t\B{t}(g)$:
$\sg_{\rm LO}(+,-)$ (lower dashed line), $\sg_{2+3}(+,-)$
 (lower solid),
$\sg_{\rm LO}(+,+)$ (upper dashed) and $\sg_{2+3}(+,+)$
 (upper solid)
for $1<\protect\sqrt{s}/2m<1.4$: {\bf (a)} $\theta_c=0$;
{\bf (b)} $\theta_c=30^0$.
}
\end{figure}
Fig.\ 7(a) gives $\sigma_{2+3}$ and $\sigma_{\rm LO}$  for $t$-quark production
in the range $1<\sqrt{s}/2m<1.4$ for $J_z=0$ and
$J_z=\pm 2$, without angular
cuts. Fig.\ 7(b) is the same except with $\theta_c=30^0$.
We notice that the angular cuts do not make a big difference.
This is because there is no
peaked behaviour in the forward/backward directions
as for $b\bar{b}$ production.
As explained earlier, the (near) absence of resolved contributions
makes the angular cuts less important experimentally as well.
The most interesting feature
of the HOC is that just above threshold, the HOC to $\sigma(+,+)$
 completely
dominate. There is no similar behaviour from $\sigma(+,-)$.
This suggests that the $J_z=0$ channel is ideal for maximizing
the top cross section not too far above threshold. At any rate,
this drastic spin dependence of the HOC is of theoretical
interest by itself and could be tested at the $b\bar{b}$
threshold as well. As the cross section is actually a function
of only $\sqrt{s}/m$ (or $\beta$) and $\alpha_s(Q^2)$ (times an
overall  factor of $e_Q^4/s$), the only
difference would be an increase in the HOC for $b\bar{b}$ relative
to its  corresponding LO term, due to an increase
in $\alpha_s$. In fact, the only ambiguity in the predictions
is the choice of scale $Q^2$
in $\alpha_s(Q^2)$. Varying
$Q^2$ in the range $s/4 < Q^2 < 4s$, for
$\sqrt{s} = 400 \,\, \mbox{\rm GeV}$, gives $\alpha_s$ in the range
$0.0878 < \alpha_s < 0.104$ and a corresponding variation in the
magnitude of the corrections.

\begin{figure}
\hspace{1.25in}\noindent
\epsfig{figure=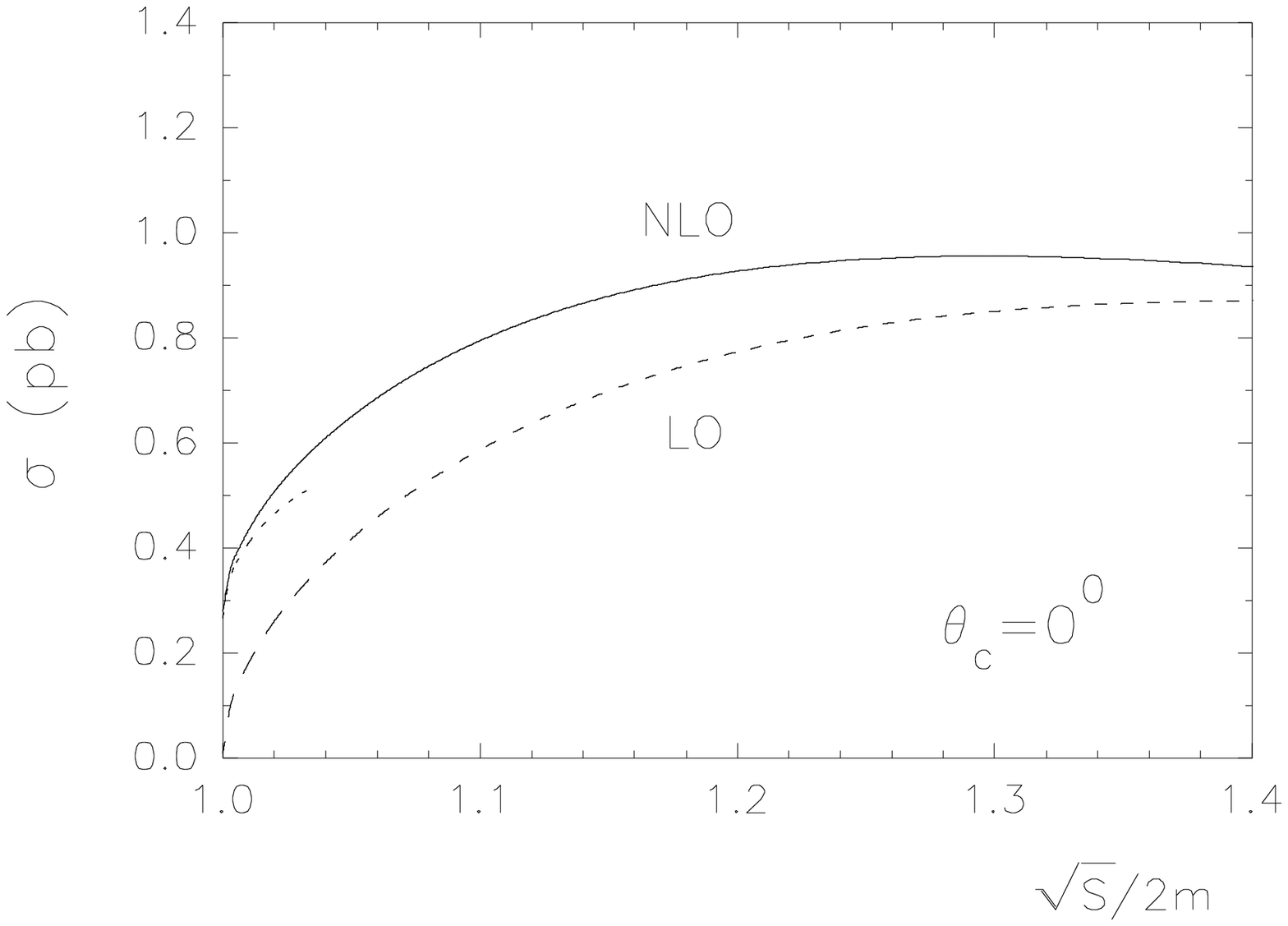,width=3.6in,height=3.6in,rheight=0in,rwidth=0in}
\vspace{5.9in}
\vspace{-2.1in}
\caption{ \protect\small\baselineskip=12pt
Unpolarized cross sections corresponding to
Fig.\ 7(a): $\sg_{2+3}$ (solid line), $\sg_{\rm LO}$
(dashed) and the small $\beta$ approximation (dotted).
}
\end{figure}
Fig.\ 8 gives the unpolarized cross sections corresponding
to Fig.\ 7(a).
We also plot the small $\beta$ (threshold region) approximation
of Ref.\ \ref{Drees}. Our results agree with this approximation
just above threshold. We see that the approximation
breaks down for $\sqrt{s}/2m \mbox{\raisebox{-.6ex}{$\,
\stackrel{\textstyle >}{\sim}\,$}} 1.02$.
As expected, we found that almost all of the correction
comes from $\sigma_{\rm S}$, i.e. $\sigma_{\rm H}$ is almost
negligible not too far above threshold. We found the same was
true for $\Delta\sigma_{\rm S}$, $\Delta\sigma_{\rm H}$.

\newpage
\vglue 1cm
\begin{center}\begin{large}\begin{bf}
VII. CONCLUSIONS
\end{bf}\end{large}\end{center}
\vglue .3cm

We have obtained complete analytical results for the production
of heavy-quark pairs by polarized and unpolarized photons in
next-to-leading order. Using these expressions, we computed
cross sections for $b$- and $t$-quark  production by photons
having net spin $J_z=0,\pm 2$. From the $b\bar{b}$ cross sections,
we determined the background to $\gamma\gamma\rightarrow H^* \rightarrow
b\bar{b}$
(standard model)
coming from $\gamma\gamma\rightarrow b\bar{b}(g)$ (2-jet) for
$<\lambda_1\lambda_2>
= 0.8$. The HOC to the $J_z=0$ channel were found to be
large for $s\gg 4m^2$.
For the experimental setup considered, the background
was safely below the Higgs signal (but still sizable)
for $90\mbox{\raisebox{-.6ex}{$\,\stackrel{\textstyle <}{\sim}\,$}}
{}m_{\rm H}\mbox{\raisebox{-.6ex}{$\,\stackrel{\textstyle <}{\sim}\,$}}
150\,\,\mbox{\rm GeV}$, even
after inclusion of HOC. For $t$-quark production, not too
far above threshold, the dominant contribution came from the
$J_z=0$ channel. Just above threshold, the HOC to this channel
completely dominate.

\vglue 1cm
\begin{center}\begin{large}\begin{bf}
ACKNOWLEDGMENTS
\end{bf}\end{large}\end{center}
\vglue .3cm
We would like to thank D.\ Atwood, G.\ Jikia and
O.\ Teryaev for discussions and J.\ Peralta
for help in the early stages.
This work was also supported by the
Natural Sciences and Engineering Research Council of Canada and
by the Quebec Department of Education.

%\vglue 1cm
\begin{center}\begin{large}\begin{bf}
APPENDIX A
\end{bf}\end{large}\end{center}
\vglue .3cm

\setcounter{equation}{0}
\renewcommand{\theequation}{A\arabic{equation}}

Here we present
the momentum parametrizations in the frame where $p_4$ and $k$
are back-to-back. We find
\begin{eqnarray}
\nonumber
p_1 &=& (\omega_1; 0, \cdots, |{\bf p}|\sin\psi,
|{\bf p}|\cos\psi - \omega_2), \\ \nonumber
p_2 &=& (\omega_2; 0, \cdots, 0, \omega_2), \\ \nonumber
k &=& (\omega_k;\, \cdots, \omega_k\sin\theta_1\cos\theta_2,
\omega_k\cos\theta_1), \\ \nonumber
p_4 &=& (E_4;\, \cdots, -\omega_k\sin\theta_1\cos\theta_2,
-\omega_k\cos\theta_1), \\
p_3 &=& (E_3; 0, \cdots, |{\bf p}|\sin\psi,
|{\bf p}|\cos\psi),
\end{eqnarray}
where
\begin{eqnarray}
\nonumber
\omega_1 &=& \frac{s+t}{2\sqrt{S_2}}, {\rm \hspace{.2in}}
\omega_2 = \frac{s+u}{2\sqrt{S_2}}, {\rm \hspace{.2in}}
\omega_k = \frac{s_2}{2\sqrt{S_2}}, {\rm \hspace{.2in}}
E_4 = \frac{s_2+2m^2}{2\sqrt{S_2}} \\
E_3 &=& - \frac{T+U}{2\sqrt{S_2}}, {\rm \hspace{.2in}}
|{\bf p}| =  \frac{\bar{y}}{2\sqrt{S_2}}, {\rm \hspace{.2in}}
\cos\psi = \frac{us_2 - s(t+2m^2)}{(s+u)\bar{y}}
\end{eqnarray}
in agreement with Ref.\ \ref{Been}. For $p_1$, $p_2$, $p_3$ the dots
represent zeros. For $k$, $p_4$ they represent components which
depend on the remaining $n-4$ angles of $k$. Since these components
do not contribute to $[\Delta]|M|^2_{2\rightarrow 3}$, those angles were
trivially integrated over in the phase space (\ref{twothreeps}).

%\newpage
\vglue 1cm
\begin{center}\begin{large}\begin{bf}
APPENDIX B
\end{bf}\end{large}\end{center}
\vglue .3cm

\setcounter{equation}{0}
\renewcommand{\theequation}{B\arabic{equation}}

In this appendix
we list the coefficients for the various cross sections.
For $\Delta d\sigma_{\rm vse}/dvdw$ given in Eq.\ \ref{vse}, the
coefficients $\Delta A_i$ are
\begin{eqnarray}
\nonumber \Delta A_1&=& 2 [-1-u/s+u^2/st+m^2s/tu-2m^2u/t^2]\\
\nonumber\Delta A_2&=&-4[4(6u/t-4s/u-t/T)m^2/t-4sT/t^2
       -16s/t+24u/s+s/T+4tu/sT]m^2/T\\
\nonumber \Delta  A_3&=&16[(7s/t+3-3t/u)m^4/t^2+(4u^2/st
          +2t/s-s^2/tu)m^2/t+t/s+u^2/st]\\
\Delta    A_4&=&4[4m^2u/t^2-3s/t+4u/s]m^2/T
\end{eqnarray}

For $d\sigma_{\rm vse}/dvdw$ given in Eq.\ \ref{vse}, the
coefficients $A_i$ are
\begin{eqnarray}
\nonumber  A_1&=& 2[u/t - 2m^2/t + sm^2/tu -4m^4/tu -4m^4/t^2] \\
\nonumber   A_2&=&4[4(12sT/t^2+t/T+6)m^4/uT+(s^2/uT
           -4s^2/tu-4s^2T/t^2u \\
\nonumber  &-&12s/t-2t/T-8)m^2/T-4] \\
\nonumber   A_3&=&16[12m^6s/t^3u+(s/t-14-8t/u)m^4/t^2-(3s/t+7+t/u)m^2/t
           +u/t] \\
    A_4&=&-4[4(2s/t+t/T)m^4/tu+(t/T-4)m^2s^2/t^2u
           +2t/T-4]
\end{eqnarray}
We note that
\begin{equation}
\label{A1LO}
[\Delta] A_1 + (t\leftrightarrow u) = (2m)^2 [\Delta] |M|_{\rm LO}^2/
(N_Ce^4e_Q^4\mu^{4\varepsilon}).
\end{equation}

For $\Delta d\sigma_{\rm box}/dvdw$ given in Eq.\ \ref{box}, the
coefficients $\Delta B_i$ are
\begin{eqnarray}
\nonumber\Delta B_1&=&\Delta A_1\\
\nonumber\Delta B_2&=&4(s+4t)m^4/stu+2(s/u+4u/s)m^2/s-s^2/tu-2s/u+
           t/u-4u/s\\
\nonumber\Delta B_3&=&12m^4/tu+2m^2(t-u)/tu-s^2/tu-t/u\\
\nonumber\Delta B_4&=&[8(t-u)/s-3s/t-s/u]m^2/s+5u/s-t/s\\
\nonumber\Delta B_5&=&4sT/tu+(t/u-1)m^2t^2/sT^2+4(s^2-2t^2)/st-
7tu/sT+t^2(s-3t)/sTu\\
\Delta B_6&=&2(s/t-4)m^4/tu-2m^2(t-u)/tu+s^2/tu+t/u\\
\nonumber\Delta B_7&=&4(u-4t)m^4/t^2u+2m^2(u-t)/tu+s^2/tu+t/u; {\rm
\hspace{.2in}}
\Delta B_8=(1+t^2/Ts)(t/u-1)
\end{eqnarray}

For $d\sigma_{\rm box}/dvdw$ given in Eq.\ \ref{box}, the
coefficients $B_i$ are
\begin{eqnarray}
\nonumber B_1&=&A_1; {\rm \hspace{.2in}}
B_2=(2m^2-s)[(2m^2-u)/st-2/u]-2m^2(6m^2+t)/su \\
\nonumber B_3&=&-4m^4/tu+2m^2(u-t)/tu+s^2/tu+t/u; {\rm \hspace{.2in}}
B_4=-2m^2s/tu+2\\
\nonumber B_5&=&2[(2m^2+s)(2s/t+t/T)-t^3/T^2]/u; {\rm \hspace{.2in}}
B_6=8m^4/tu+2m^2/u-s^2/tu-t/u \\
B_7&=&-4m^4/tu+2m^2(t-3u)/tu-s^2/tu-t/u; {\rm \hspace{.2in}}
B_8=2(s+t^2/T)/u
\end{eqnarray}

For $\Delta |M|_{2\rightarrow3}^2$ given in Eq.\ \ref{M23} and
 $\Delta d\sigma_{\rm Br}/dvdw$ given in Eq.\ \ref{br}, the
coefficients $\Delta e_i$ are
\begin{eqnarray}
\nonumber\Delta \tilde{e}_1 &=& -16(s/u-s/t-2)m^4/u -
4[s_2(2+2s/u-t/s+u/s+2t^2/s^2
   + 8tu/s^2)+2s \\
\nonumber&-& 4tu/s]m^2/u + s_2(4s/u-4-8t^2/su-5t/s) \\
\nonumber\Delta e_2&=&-4[2m^2(2/s_2u+2/s_2t+1/s^2-u/s^2t)+6/t-t/s^2+
u^2/s^2t]/u\\
\nonumber\Delta e_3&=&-2[8m^4(1/tu-1/s_2u-s/s_2t^2)-2(4s/s_2+s/t-1)m^2/t
-3s/u-5s/t \\
\nonumber&-& (2s^3/tu-su/t+3t+2u+u^2/t)/s_2] \\
\nonumber\Delta e_4&=&-2(2m^2s/t+2s+u)m^2/t; {\rm \hspace{.2in}}
\Delta e_5 = 0 \\
\nonumber\Delta \tilde{e}_6 &=& [32m^4/u -
 4m^2(t/u+5+t^2/su+5t/s+2t^3/s^2u+10t^2/s^2
  + 8tu/s^2) \\
\nonumber&+& 4st/u-16t-8t^2s_2/su-5ts_2/s]/2 ; {\rm \hspace{.2in}}
\Delta \tilde{e}_7=-2m^2 \\
\nonumber\Delta e_8&=&4m^4(s/tu+2/u+1/s)-2m^2(s^2/tu+2s_2/u-1+t/s)-
(s^2/t+s+3s_2 \\
\nonumber &+&t^2/s+3tu/s)s/u;{\rm \hspace{.2in}}
\Delta e_9=8(1/u-1/s_2)/t\\
\nonumber\Delta e_{10}&=&4[2(1/u+2s/s_2t-2/s_2)m^2/u+s_2/tu-3/s_2
-u/ts_2]\\
\nonumber\Delta \tilde{e}_{11} &=& 8m^4(s/t+t/u+s_2/s)
- 2m^2[2s^2/u+2s^2/t+s
   + s_2(2t/u+2+4u/t+t/s+u/s)]\\
\nonumber &-& (s^2+t^2)(t+u)/u; {\rm \hspace{.2in}}
\Delta \tilde{e}_{12}=m^2s; {\rm \hspace{.2in}}
\Delta e_{13}=-m^2(4m^2s/t+2s+u)\\
\Delta \tilde{e}_{14}&=&-m^4s; {\rm \hspace{.2in}}
\Delta e_{15}=4[2m^2/u-2m^2/t-t/u+u/t]/s^2 \\
\nonumber\Delta e_{16}&=&a_1+a_1(t\leftrightarrow u)
-8(m^2+m^2t/s-t)/su+2[t^2
(t/u+2)-u^2(u/t+2)]/s^2s_2
\end{eqnarray}
where
\begin{equation}
 a_1=t[4m^2(7s/t+1+3t/s+9u/s)-2s+6su/t+4t+u]/s_2us
\end{equation}

For $|M|_{2\rightarrow3}^2$ given in Eq.\ \ref{M23} and
$d\sigma_{\rm Br}/dvdw$ given in Eq.\ \ref{br}, the
coefficients $e_i$ are
\begin{eqnarray}
\nonumber\tilde{e}_1 &=& 2[16m^6(t+u)/tu + 16m^4(s_2/u+s/t+3)
  +2m^2(25s_2-2t+2s_2^2/u) \\
\nonumber &+& s_2(2s+5t+10u)]/u; {\rm \hspace{.2in}}
e_2=-8[2m^2(t+u)/s_2t+3]/tu \\
\nonumber e_3&=&-2[16m^6(t+u)/s_2tu+8(4+u/t-2s/s_2)m^4/u
 + 2m^2 (6t/u+6-u/t-s/t+s/s_2) \\
\nonumber &-& 2s^2/u-3st/u-2s-s_2+3st/s_2-2t^3/s_2u]/t \\
\nonumber e_4&=&2m^2(2m^2/t+1)^2; {\rm \hspace{.2in}} e_5=-4m^2/u \\
\nonumber \tilde{e}_6 &=& -16m^4/u+2m^2(5s/u+7t/u+11) + t(4s/t-s/u+t/u+17);
{\rm \hspace{.2in}}
\tilde{e}_7=2m^2\\
\nonumber e_8&=&8m^6(s/t+2+u/s)/su+4(3s/t+4+2s_2/s)m^4/u+2m^2(
                s_2/u-2-t/s) \\
\nonumber &-&(s^2/t+s+3s_2+t^2/s+3tu/s)s/u; {\rm \hspace{.2in}}
e_9=8(s_2+u)/ts_2u \\
\nonumber e_{10}&=&-4[8m^4/u+2m^2(s/u+2)-s_2^2/u+4s+t+3u]/ts_2\\
\nonumber \tilde{e}_{11} &=& 16m^6[(t+u)^2/tu+s_2/s]/s +
8m^4[(s-t)/u+t^2/su
  + 3(t+u)/s + u^2/st] \\
\nonumber  &+& 2m^2[u(s+u)/t + t(t+u)/s -ts_2/u
  - 7s_2 - 2ts_2/s - us_2/s] + (s^2+t^2)(t+u)/u \\
\nonumber \tilde{e}_{12}&=&m^2(8m^4/s-4m^2-s); {\rm \hspace{.2in}}
e_{13}=m^2(8m^4/t+4m^2-t); {\rm \hspace{.2in}}
\tilde{e}_{14}=2m^6; {\rm \hspace{.2in}}
e_{15}=0 \\
e_{16}&=&-2[22m^2(t+u)/tu+3t/u+14+u/t]/s_2
\end{eqnarray}

\vglue 1cm
\begin{center}\begin{large}\begin{bf}
APPENDIX C
\end{bf}\end{large}\end{center}
\vglue .3cm

\setcounter{equation}{0}
\renewcommand{\theequation}{C\arabic{equation}}

We give here the bremsstrahlung integrals, $I_i$, appearing in
Eq.\ (\ref{br}).
They are defined as
\begin{equation}
I_i = \frac{1}{2\pi} \int\,\, d\Omega f_i\,\,; {\rm \hspace{.4in}}
(2m)^2 [\Delta] |M|^2_{2\rightarrow 3} \equiv C \sum_i [\Delta]
e_i f_i/s_2^{n_i}
\end{equation}
(see (\ref{M23})).
The $f_i$ may be explicitly expressed as functions of
$\theta_1$ and $\theta_2$ using the expressions in
Appendix A. All the integrals here are 4-dimensional
(i.e.\ $\varepsilon=0$ in (\ref{angint})) and are determined
using the general forms given in Ref.\ \ref{Been}.

First we list the four basic integrals,
\begin{eqnarray}
\nonumber I_6 &=& \frac{2 S_2}{s_2 \bar{y}} \ln\frac{T+U-\bar{y}}
{T+U+\bar{y}},
{\rm \hspace{.4in}} I_6' \equiv \frac{2S_2}{s_2(s+t)}\ln\frac{S_2}
{m^2} \\
\nonumber I_8 &=& \frac{4S_2}{s_2\sqrt{s}} \frac{1}{\sqrt{x_8}}
\ln\frac{x_8+s_2^2s+2s_2\sqrt{sx_8}}{x_8+s_2^2s-2s_2\sqrt{sx_8}},
{\rm \hspace{.2in}} x_8 \equiv 4m^2(s_2s+tu) + s_2^2s \\
 I_{11} &=& \frac{4S_2}{s_2\sqrt{stx_{11}}}
\ln\frac{x_{11}+st-2\sqrt{stx_{11}}}{x_{11}+st+2\sqrt{stx_{11}}},
{\rm \hspace{.2in}} x_{11} \equiv 4m^2(s_2-t) + st
\end{eqnarray}
Define,
\begin{equation}
z_1 \equiv 2m^2s+s_2s-tu, {\rm \hspace{.2in}} z_2 \equiv s_2u-2m^2s-st,
{\rm \hspace{.2in}}
z_3 \equiv m^2s-tu, {\rm \hspace{.2in}} z_4 \equiv 2m^2s-tu,
{\rm \hspace{.2in}} z_5 \equiv
2m^2+t
\end{equation}
We may now express the remaining integrals in terms of those listed
above:
\begin{eqnarray}
\nonumber I_5 &=& - \frac{2S_2z_4}{m^2(s+t)^3}
+ \frac{I_6'z_1}{(s+t)^2} \\
\nonumber I_9 &=& \frac{1}{4S_2(s+t)^3} \{ 2z_1(s_2-t)(s+t)s_2
- (z_1^2+2z_3S_2s)(2m^2+s_2)\} + \frac{I_6'}{4(s+t)^2}
(z_4^2+2m^2sz_3) \\
\nonumber I_{10} &=& \frac{z_1}{(s+t)^2} - \frac{I_6'z_4}{2(s+t)};
{\rm \hspace{.2in}}
I_{12} = \frac{8S_2}{m^2s_2t} \left(\frac{z_5}{x_{11}}-\frac{1}{s_2}
\right)
- 2I_{11}\frac{z_5(s_2-t)}{x_{11}t} \\
\nonumber I_{13} &=& -\frac{8S_2z_5}{x_{11}(s_2-t)m^2t}
- 2\frac{I_{11}}{t} \left(1-\frac{s_2z_5}{x_{11}}\right) \\
\nonumber I_{14} &=& \frac{16S_2}{stx_{11}} \left(\frac{12z_3}{x_{11}t}
+ \frac{\bar{y}^2+s_2^2}{s_2^2m^2}\right) - \frac{4I_{11}}{x_{11}st}
\{s_2u-sz_5-3z_5(s_2-t)\frac{s}{t} \left(1-\frac{s_2z_5}{x_{11}}\right)\}
\\
\nonumber I_{15} &=& \frac{s_2}{4S_2\bar{y}^4} \{ (z_2^2 + 2S_2sz_3)(U+T)
+ 2\bar{y}^2z_2(s_2-t)\} + \frac{I_6s_2^2}{4\bar{y}^4}
\{ (z_4-u^2)^2 + 2m^2sz_3 \} \\
{}I_{16} &=& \frac{z_2}{\bar{y}^2} - \frac{I_6s_2(z_4-u^2)}{2\bar{y}^2}
\end{eqnarray}
The integrals were put into the above form using REDUCE.
The integrals not listed here (including the $n$-dimensional ones
not given in Ref.\ \ref{Been})
are straightforward and have
been substituted directly in (\ref{br}).
As an aside, we point out that $x_{11}(t\leftrightarrow u)$ vanishes for
$v=1/2$, $w=w_1$. Hence one must avoid reaching {\em exactly}
the lower bound (as for the upper) of the $w$ integral, in numerical
calculations.

%\newpage
\vglue 1cm
\begin{center}\begin{large}\begin{bf}
REFERENCES
\end{bf}\end{large}\end{center}
\vglue .3cm

   \begin{list}{\arabic{enumi}.}
    {\usecounter{enumi} \setlength{\parsep}{0pt}
     \setlength{\itemsep}{3pt} \settowidth{\labelwidth}{99.}
     \sloppy}
\item [*]
On leave from High
Energy Physics Institute, Tbilisi State University, Tbilisi,
Republic of Georgia.
\item \label{Nason}
P.\ Nason, S.\ Dawson, and R.K.\ Ellis, Nucl.\ Phys.\ {\bf B303},
607 (1988); {\bf B327}, 49 (1989) and erratum {\bf B335}, 260
(1990).
\item \label{Been}
W.\ Beenakker, H.\ Kuijf, W.L.\ van Neerven, and J.\ Smith, Phys.\
Rev.\ {\bf D40}, 54 (1989).
\item \label{Drees}
J.\ Smith and W.L.\ van Neerven, Nucl.\ Phys.\ {\bf B374}, 36 (1992);
J.H.\ K\"{u}hn, E.\ Mirkes, and J.\ Steegborn, Z.\ Phys.\ {\bf C57},
615 (1993); M.\ Drees, M.\ Kr\"{a}mer, J.\ Zunft, and P.M.\ Zerwas,
Phys.\ Lett. {\bf 306B}, 371 (1993).
\item \label{Hab}
J.F.\ Gunion and H.E.\ Haber, Phys.\ Rev.\ {\bf D48}, 5109 (1993).
\item \label{Borden}
D.L.\ Borden, V.A.\ Khoze, J.\ Ohnemus, and W.J.\ Stirling,
Phys.\ Rev.\ {\bf D50}, 4499 (1994).
\item \label{Seigel}
W.\ Siegel, Phys.\ Lett.\ {\bf 84B}, 193 (1979).
\item \label{ward}
J.C.\ Ward, Phys.\ Rev.\ {\bf 78}, 182 (1950).
\item \label{DY}
A.P.\ Contogouris, B.\ Kamal, and Z.\ Merebashvili, Phys.\ Lett.\
{\bf 337B}, 169 (1994).
\item \label{LLbar}
A.P.\ Contogouris, O.\ Korakianitis, F.\ Lebessis, and Z.\
Merebashvili, Phys.\ Lett.\ {\bf B344}, 370 (1995).
\item \label{Korner}
G.A.\ Schuler, S.\ Sakakibara, and J.G.\ K\"{o}rner,
Phys.\ Lett.\ {\bf 194B}, 125 (1987); J.G.\ K\"{o}rner and
M.M.\ Tung, Z.\ Phys.\ {\bf C64}, 255 (1994).
\item \label{Jones}
I.\ Jack, D.R.T.\ Jones and K.L.\ Roberts, Z.\ Phys.\ {\bf C62},
161 (1994); {\bf C63}, 151 (1994); M.\ Ciuchini,
E.\ Franco, L.\ Reina, and L.\ Silvestrini, Nucl.\ Phys.\ {\bf B421},
41 (1994).
\item \label{PV}
G.\ Passarino and M.\ Veltman, Nucl.\ Phys.\ {\bf B160}, 151 (1979).
\item \label{Verm}
J.\ Vermaseren, FORM User's Manual (CAN, Amsterdam, 1991).
\item \label{Hearn}
A.C. Hearn, REDUCE User's Manual Version 3.3
(Rand Corporation, Santa Monica, CA, 1987).
\item \label{EllFurm}
R.K.\ Ellis, M.A.\ Furman, H.E. Haber, and I.\ Hinchliffe,
Nucl.\ Phys.\ {\bf B173}, 397 (1980).
\item \label{Abe}
CDF Collaboration, F.\ Abe et al., Phys.\ Rev.\ Lett.\ {\bf 73},
225 (1994);  Phys.\ Rev.\ {\bf D50}, 2966 (1994).
\end{list}

\newpage

\pagestyle{empty}

\begin{center} \begin{Large} \begin{bf}

Erratum: Heavy-quark production by polarized and \\ \mbox{} \\
\vspace{-.5cm}
unpolarized photons in next-to-leading order \\ \mbox{} \\
\vspace{-.5cm}
\mbox{[Phys.\ Rev.\ D 51, 4808 (1995)]}

\end{bf} \end{Large} \end{center}
%\vglue 0.35cm
{\begin{center}
 B.\ Kamal, Z.\ Merebashvili and A.P.\ Contogouris \end{center}}

\vglue .2cm

\noindent
\hspace{1in} PACS number(s): 13.65.+i, 13.88.+e, 14.65.-q, 14.70.Bh

\vglue .5cm

Due to a trivial algebraic error, Eq.\ (30) is incorrect. On the fourth
line one should make the replacement $-1 \rightarrow 2\ln x$. This gives
for Eq. (30)
\vspace{.2cm}
\begin{eqnarray*}
 \frac{d\sigma_{\rm Br}}{dvdw} &=& \cdots
+ 8\pi\alpha_s C_F C_{\varepsilon} \frac{d\sigma_{\rm LO}}{dvdw}
\frac{1}{s\beta}
  \left((2m^2-s)  \left\{2\ln x\left(2 \ln \frac{sv}{m^2}
-\frac{1}{\varepsilon} + 1\right) \right. \right.
 \\  &  & \mbox{} 
\\ 
&-& \left. \left. 2\left[{\rm Li}_2\left(
\frac{-4\beta}{(1-\beta)^2}\right)
+ \ln^2 x\right]\right\} 
+ 2s\beta\left[1-2\ln\frac{sv}{m^2} + \frac{1}{\varepsilon}\right]
 \right).
\end{eqnarray*}

\vspace{.2cm}
The essential conclusions remain unchanged as the features of the 
figures are the same. Various checks were done to ensure that the
analytical results (not completely presented in any other works) are 
now correct. We now agree exactly with Table 1 of K\"{u}hn et.\
al.\ [3] and the corresponding polarized results in Table 1 of
[G.\ Jikia and A.\ Tkabladze, Phys.\ Rev.\ D {\bf 54}, 2030 (1996)].
We also agree exactly with the unpolarized squared amplitude of
[J.F.\ Gunion and Z.\ Kunszt, Phys.\ Lett.\ B {\bf 178}, 296 (1986)]
(although our 3-jet cross sections had a minor numerical error which
made the ones presented in Figures 4 and 5 a bit too large on average) and all
the bremsstrahlung integrals, including the $n$-dimensional ones, have
been checked numerically. Finally, we agree exactly with the unpolarized
virtual+soft cross section of [2] and consequently obtain the correct
threshold behavior. The soft part of [2] may be obtained by making
the substitution $\ln(sv/m^2) \rightarrow \ln \delta$ in the above
expression, as may be easily derived.

\end{document}